\documentclass{aa}
\usepackage{natbib}
\bibpunct{(}{)}{;}{a}{}{,}
\usepackage{graphicx}

\begin{document}


\title{Baryonic pollution in gamma-ray bursts:\\
 the case of a magnetically driven wind emitted\\
from a disk orbiting a stellar mass black hole}
\author{Fr\'ed\'eric Daigne\inst{1} \and Robert Mochkovitch\inst{2}}
\institute{Max-Planck-Institut f\"ur Astrophysik,
Karl-Schwarzschild-Str 1. Postfach 1317 D-85741 Garching bei M\"unchen\\
Present address: Service d'Astrophysique, CEA/Saclay, F-91191 Gif sur Yvette Cedex
\and Institut d'Astrophysique de Paris, 98 bis bd. Arago F-75014 Paris}
\mail{daigne@discovery.saclay.cea.fr}
\date{Received 15 October 2001 / Accepted 25 February 2002}
\abstract{Most models for the central engine of gamma-ray bursts involve a stellar mass
black hole surrounded by a thick disk formed after the merging of a 
system of compact objects or the collapse of a massive star.
Energy released from the accretion of disk material by the black hole or
from the rotation of the hole itself extracted by the Blandford-Znajek
mechanism powers a relativistic wind along the system axis.    
Lorentz factors of several hundreds are needed to solve the compactness 
problem in the wind which implies the injection 
of a tremendous
power into a very small 
amount of
matter. The Blandford-Znajek mechanism, where the outflow follows magnetic 
field lines anchored to the black hole is probably the best way to prevent 
baryonic pollution and can even initially produce a purely leptonic wind.
In this paper we rather study the wind emitted from the inner part of the
disk where
the risk of baryonic pollution is much larger since the outflow originates 
from high density regions. We show that the  
baryonic load of this wind sensitively depends on the disk temperature and 
magnetic field geometry and that the outflow can become ultra-relativistic 
(Lorentz 
factor $\Gamma>100$) under quite restrictive conditions only.
Conversely, if $\Gamma$ remains of the order of unity the dense wind emitted 
from the 
inner disk could help to confine the central jet but may also represent a 
source of baryon contamination for the Blandford-Znajek
mechanism.
}
\titlerunning{Baryonic pollution in GRBs: the case of a magnetically driven wind}
\authorrunning{F. Daigne \& R. Mochkovitch}
\maketitle
\section{Introduction}
The discovery of the first optical counterparts to gamma-ray bursts 
(hereafter GRBs) in 1997 \citep{vanparadijs:97} has shown that most (if not all) 
GRBs are
located at cosmological distances. About 20 redshifts have now been 
measured from
$z=0.43$ for GRB 990712 to $z=4.5$ for GRB 000131 (with however the
peculiar case of GRB 980425 which appears to be associated to a nearby type Ic 
supernova at $z=0.01$; \cite{galama:98}). The energy radiated by these cosmological GRBs 
in the BATSE range (20 -- 1000~keV)   
goes from $5\,10^{51}$ ergs for GRB 970228 and GRB 980613 (at $z=0.695$ and
1.096) to $2\,10^{54}$ ergs for 
GRB 990123 (at $z=1.6$) assuming isotropic emission. 
After correction for beaming, the true energy output appears to be less
scattered, clustered around $E_{\gamma}\sim 5\,10^{50}$ ergs
\citep{djorgovski:01}.   
Among the sources which have been proposed 
to explain such a huge
release of energy in a time of seconds, the most popular are mergers 
of compact
objects (neutron star binaries or neutron star -- 
black hole systems) \citep{narayan:92,meszaros:92a,mochkovitch:93}
or massive stars which collapse to a black hole (collapsars) 
\citep{woosley:93,paczynski:98}. In all these cases,
the resulting configuration is a stellar mass black hole surrounded by a thick
torus made of stellar debris or of infalling stellar material
partially supported by centrifugal forces. The location of the detected 
optical counterparts well inside their host galaxies and often associated
with regions of star formation appears to favor the collapsar scenario 
\citep{paczynski:98,owens:98,klose:00}. 
Double neutron star or neutron star -- black hole mergers should 
generally be 
observed at the periphery of the host galaxy due to the long delay 
before coalescence and the large velocity imparted to these systems 
by two successive supernova explosions.
They can however still be invoked in the case of shorts bursts, 
for which no optical 
counterpart has been detected.\\

If black hole
+ thick disk configurations are indeed at the origin of GRBs, the released
energy will ultimately come from the accretion of disk material by the black 
hole or from the rotational energy of the hole itself extracted by the
Blandford-Znajek mechanism \citep{blandford:77}. In a first step, the energy must be injected into a
relativistic wind whose existence has been directly inferred from the 
observations of radio scintillation in GRB 970508 \citep{waxman:98}
and which is also 
needed to avoid photon-photon annihilation and the resulting compactness 
problem \citep{baring:97}. The second step
consists of the conversion of a fraction of the wind kinetic energy
into gamma-rays via the formation of shocks, probably inside the wind itself \citep{rees:94}. 
These internal shocks can be
expected if the source generate a highly non uniform distribution of  
Lorentz factor so that rapid layers of the wind will catch up with slower ones 
at large relative
velocities. In the last step, 
the wind is decelerated when it interacts with the interstellar or 
circumstellar medium 
and the resulting (external) shock is responsible for the afterglow
observed in the X-ray, optical and radio bands \citep{meszaros:97}.\\

The physics of the afterglow is probably the
best understood since most afterglow properties can  
be interpreted in terms of solutions of the relativistic Sedov 
problem \citep{blandford:76} with synchrotron emission behind the shock \citep{sari:98}.
Models taking into account the beam geometry of the flow \citep{rhoads:97}
or different kinds of burst
environments (constant density medium or stellar wind; \citet{chevalier:00}) can be constructed 
to explain observed breaks in the lightcurves or the evolution of 
some radio afterglows.  \\

More problems remain concerning the generation of gamma-rays in the relativistic
wind. Instead of internal shocks, gamma-rays can also be emitted 
during the early 
evolution of the external
shock with however the difficulty to explain in this case the highly variable 
temporal profiles of observed bursts (\cite{sari:98} see however 
\cite{dermer:99}). 
Models of bursts produced by internal shocks are found to be 
in reasonable agreement with the observations \citep{kobayashi:97,daigne:98,daigne:00}
even if the process which produces the gamma-rays 
-- synchrotron radiation or/and comptonization -- 
remains uncertain due to problems encountered in fitting the low energy part 
of the spectra \citep{preece:98}. \\

The origin of the relativistic wind is the more complex of the 
three steps. Several proposals have been made to explain
the generation of this wind but  
few detailed calculations have been performed. 
If the burst energy comes from matter accretion  
by the black hole, the annihilation of neutrino-antineutrino pairs 
emitted by the hot disk could be a way to inject energy along the system axis,
in a region which can be expected to be depleted in baryons due to the
effect of centrifugal forces \citep{meszaros:92b,mochkovitch:93}. 
The low efficiency of this process
requires 
high neutrino luminosities and therefore high accretion rates.
In the merger case this can be achieved for short accretion timescales \citep{ruffert:99} and may explain short bursts. Conversely in the collapsar scenario, the larger mass reservoir allows the system
to maintain high accretion rates for a longer time and could then also produce long bursts
\citep{macfadyen:99}.

Another possibility is to suppose that disk energy
is extracted by a magnetic field amplified by differential
rotation up to very 
large values ($B\ga 10^{15}$ G). A magnetically driven wind
could then be emitted from the disk with a fraction of the Poynting flux
being eventually transferred to matter \citep{blandford:82}. 
Such a mechanism operates in many
classes of astrophysical objects (from T Tauri stars to AGN) but it is far
from clear that it can work in the context of GRBs where final Lorentz factors
of several hundreds are required. In a different version of the same idea,
an early conversion of magnetic into thermal energy could occur through
the reconnection of field lines above the plane of the disk in a region of
rather low density \citep{narayan:92}. 
 
An alternative
to accretion is to directly extract the rotational energy of the
black hole via the Blandford-Znajek mechanism. The available power then depends
on the rotation parameter $a$ of the hole and on the intensity of the
magnetic field pervading the horizon. If $B\ga 10^{15}$ G and 
$a \sim 1$, the power available from the Blandford-Znajek mechanism 
can be larger than $10^{52}$ erg.s$^{-1}$ with a very limited contamination by
baryons at the source since 
the field lines which guide the outflow are anchored to the 
black hole. \\

In this paper we rather concentrate on the
wind which is emitted from the inner disk. 
We want to identify the key parameters which control its
baryonic load and check whether it can reach large 
Lorentz factors or remains non relativistic. 
Our approach will be
oversimplified in comparison 
to the complexity of the real problem
so that our conclusions have to be considered as indicative only. 
In Sect.~\ref{sec:Structure} we briefly discuss the structure of the disk + black hole
configurations which are produced by NS + NS or NS + BH mergers and 
in the 
collapsar scenario. We write in Sect.~\ref{sec:Dynamics} the equations which govern the wind
dynamics from the disk to the sonic point. They are solved in Sect.~\ref{sec:MDot}
to obtain the mass loss rate and 
the terminal Lorentz factor of the wind is estimated in Sect.~\ref{sec:Gamma}. 
Our results are 
discussed in Sect.~\ref{sec:Discussion} which is also the conclusion.    
\vspace*{-2ex}

\section{The structure of the disk}
\label{sec:Structure}
\begin{figure}
\resizebox{0.75\hsize}{!}{\includegraphics{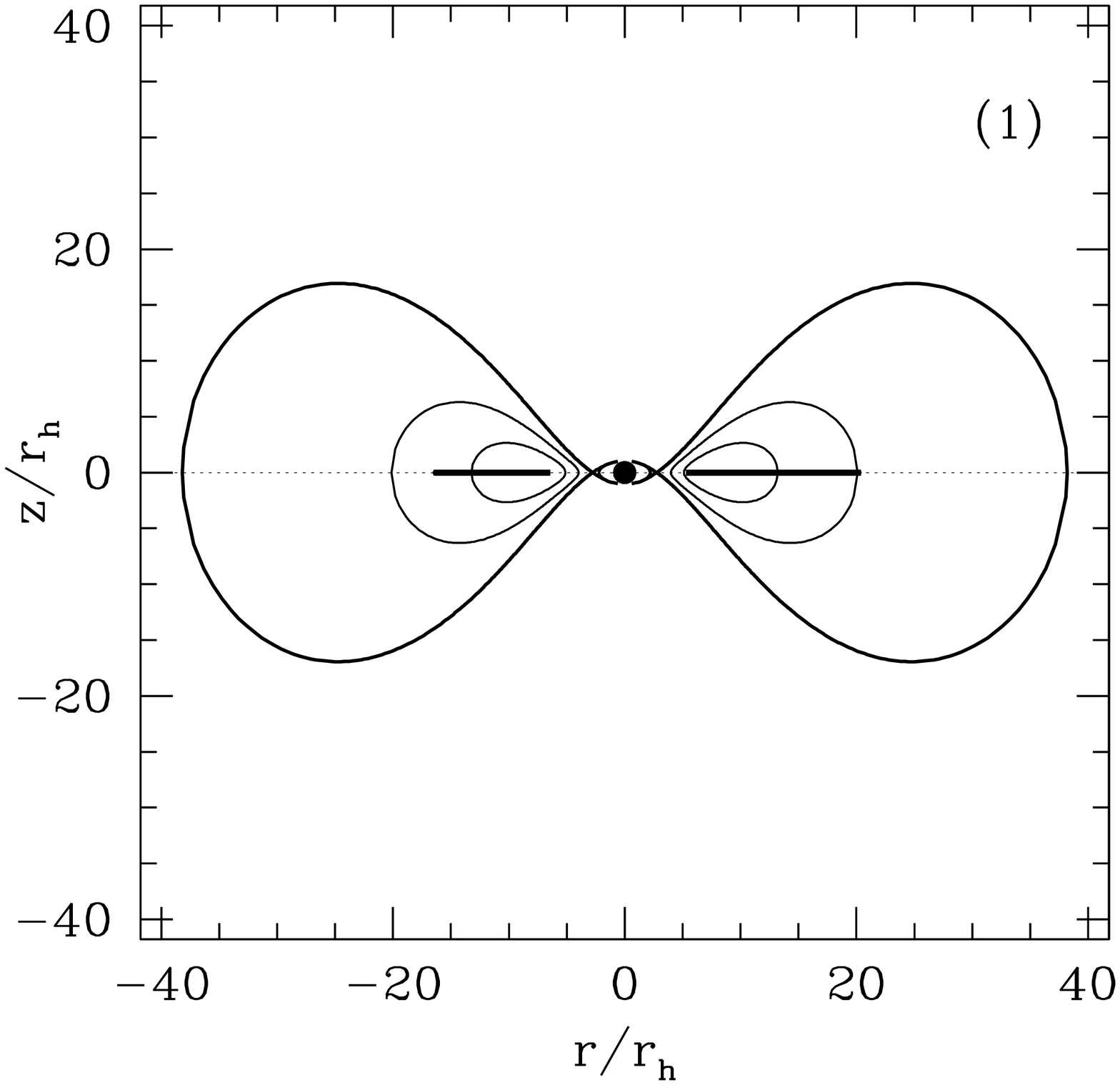}}
\resizebox{0.75\hsize}{!}{\includegraphics{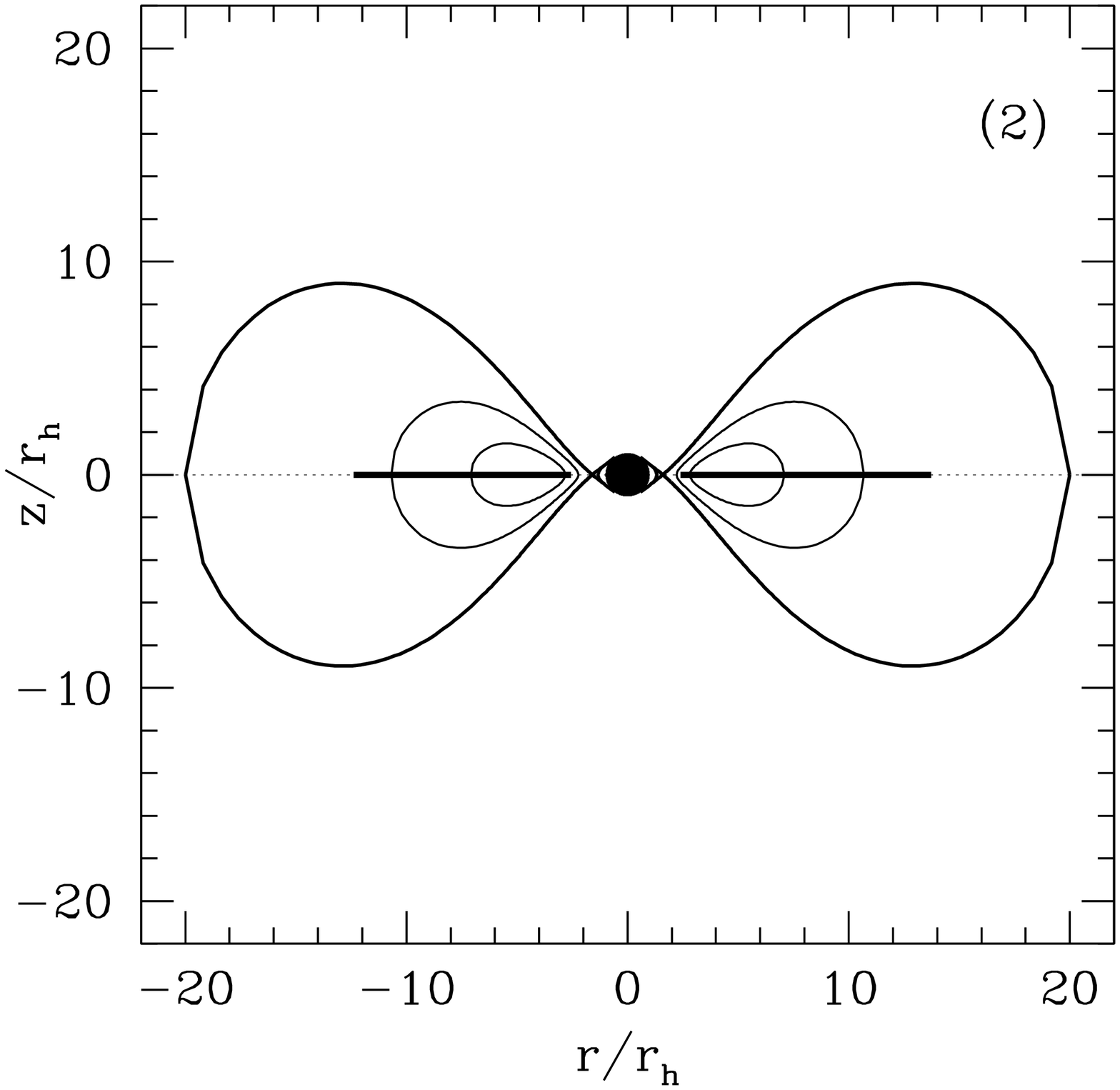}}
\resizebox{0.75\hsize}{!}{\includegraphics{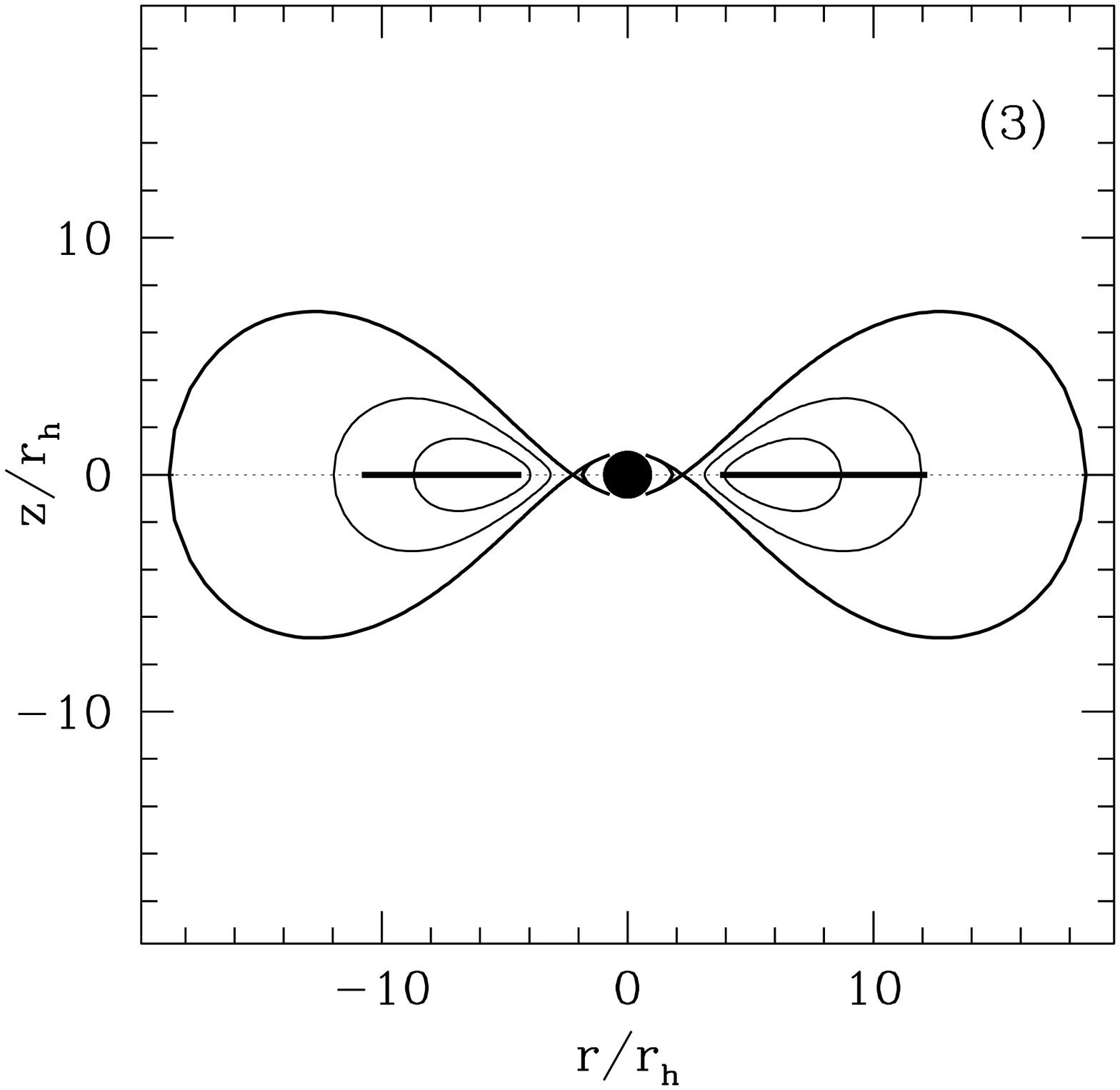}}
\caption{Equilibrium models for thick accreting disks or\-bi\-ting 
stellar mass black
holes. The disks are supported by the pressure of relativistic degenerate
electrons and the specific angular momentum is increasing outwards as
$r^{0.2}$; case 1) Schwarzschild black hole of mass $M_\mathrm{BH}=2.5\ \mathrm{M}_{\sun}$
and disk mass $M_\mathrm{D}=0.3\ \mathrm{M}_{\sun}$; 2) same $M_\mathrm{BH}$ and 
$M_\mathrm{D}$ and Kerr black hole with $a=0.8$; 3) $M_\mathrm{BH}=5\ \mathrm{M}_{\sun}$,
$M_\mathrm{D}=0.5\ \mathrm{M}_{\sun}$ and $a=0.4$. In each model thick lines in the
equatorial plane indicate the extension of the zone which is optically thick 
to neutrinos for $T_{\nu}=2$ (left) and 3 (right) MeV.}
\label{fig:Disks}
\end{figure}

Before writing the wind equations in the next section we first describe
the black hole + disk configurations which are obtained in the case of the 
three most discussed GRB sources: NS + NS or BH + NS mergers and collapsars.
We want to compare the black hole and disk masses, the hole rotation
parameter $a$, estimate which fraction of the disk is optically thick to
neutrinos and obtain the disk temperature.\\
\subsection{Mergers :}
The coalescence of neutron stars has been studied by \mbox{several} groups  
mostly in the Newtonian and post-Newtonian approximation \citep{davies:94,ruffert:96,calder:99,rosswog:00,faber:00}
The resulting
merged object obtained from two neutron stars of $1.4\ \mathrm{M}_{\sun}$ 
is made of a dense central core of $\sim 2.5\ \mathrm{M}_{\sun}$ 
in quasi-uniform rotation 
surrounded by a differentially rotating disk of $\sim 0.3\ \mathrm{M}_{\sun}$. 
In relativistic calculations, a black hole is directly 
formed during the merging event
if the total mass of the system is 30 to 70\% larger than the 
maximum rest mass of an isolated neutron star \citep{oohara:97}. 
Due to the large angular momentum of the merger the rotation parameter 
$a= Jc / {G M_\mathrm{BH}^2}$ 
of the newly formed black hole can easily exceed 0.5. \\

In the case of BH + NS mergers, the masses of the disk and black hole 
are in average larger than in NS + NS mergers. 
The hole rotation parameter depends on the fraction of neutron star
material accreted by the black hole during the merging event. \citet{janka:99}
obtain disk masses between 0.3 and $0.7\ \mathrm{M}_{\sun}$ and rotation parameter
between 0.1 and 0.5 for different assumptions regarding the relative 
masses and spins of
the neutron star and the black hole.\\

To estimate the transparency of the disks to neutrinos we have computed 
the structure of the merged object with the self-consistent field 
method originally developed
by \citet{ostriker:68} using the approach of \citet{hachisu:86} to solve
the Poisson equation. The equation of state in the disk corresponds to 
an ideal gas
of ultra-relativistic electrons and the distribution of specific 
angular momentum is $j(r)\propto r^{0.2}$ in reasonable agreement with the
results of numerical simulations \citep{ruffert:99}. We construct accreting disks 
i.e. disks with a cusp
in the equatorial plane where the total gravitational + centrifugal force 
is zero.  
If $a=0$, we model the black hole with the Paczynski-Wiita potential while 
for a Kerr black hole we use the Novikov potential \citep{artemova:96}
\begin{equation}
\Phi_\mathrm{BH}(r)=-\frac{GM_\mathrm{BH}}{(\beta-1)r_h}\left[\left(\frac{r}{r-r_h}
\right)^{\beta-1}-1\right]\ ,
\end{equation}
with
\begin{equation}
r_h=\left(1+\sqrt{1-a^2}\right)\,r_g\ \ \ \ \mathrm{and}\ \ \ \ 
\beta=\frac{r_{ms}}{r_h}-1\ ,
\end{equation}
where $r_g={GM_\mathrm{BH}}/{c^2}$ is the gravitational radius,
$r_h$ is the horizon radius and $r_{ms}$ the radius of the last stable orbit.
We have considered three cases: $M_\mathrm{BH}=2.5\ \mathrm{M}_{\sun}$, 
$M_\mathrm{D}=0.3\ \mathrm{M}_{\sun}$ with $a=0$ (case 1) and $a=0.8$ (case 2); 
$M_\mathrm{BH}=5\ \mathrm{M}_{\sun}$, 
$M_\mathrm{D}=0.5\ \mathrm{M}_{\sun}$ with $a=0.4$ (case 3). 
Cases 1 and 2 are representative of NS+NS mergers while case 3 corresponds to a BH+NS merger.
The obtained disk 
structures are
represented in Fig.~\ref{fig:Disks}. Compared to case 1, the increase of $a$ in case 2 leads 
to a denser 
($\rho_\mathrm{max}=4.2\;10^{12}$ instead of $3.3\;10^{11}$ g.cm$^{-3}$) 
and more compact disk (the radius at maximum density $r_\mathrm{max}$ 
and the external radius $r_\mathrm{ext}$ 
are 58 and 279 km in case 1, 25 and 117 km in case 2). 
In case 3, the maximum density
is $\rho_\mathrm{max}=4.3\;10^{11}$ g.cm$^{-3}$ at $r_\mathrm{max}=80.4$ km 
and the external radius is $r_\mathrm{ext}=261$ km. 
We have computed the optical thickness of these disks to
electron neutrinos in the vertical direction
\begin{equation}
\tau_{\nu_e}=\int_{-\infty}^{+\infty}\kappa_{\nu_e}\rho\;dz
\end{equation}
where 
\begin{equation}
\kappa_{\nu_e}=3.8\;10^{-19}\;\left(\frac{T_{\nu}}{1\ \mathrm{MeV}}\right)^2
\ \ \mathrm{cm^2.g^{-1}}
\end{equation}
is the neutrino opacity \citep{duncan:86}. 
The radial extension of the optically thick region
has been represented in Fig.~\ref{fig:Disks} for two temperatures, $T_{\nu}=2$ and 3 MeV. 
The optical thickness becomes larger than unity at 3 -- 5 $r_h$ and the disk remains opaque out to $r_\mathrm{out} \ga 10\ r_h$.\\

When the disk is optically thick we use the following expression 
for the neutrinosphere temperature       
\begin{equation}
T_{\nu}(r)=T_*\left(\frac{r_*}{r}\right)^{3/4}\left(
\frac{1-\sqrt{\frac{r_\mathrm{in}}{r}}}{1-\sqrt{\frac{r_\mathrm{in}}{r_*}}}\right)^{1/4}
\label{eq:TThick}
\end{equation}
where $r_\mathrm{in}$ is the disk internal radius and $T_*$ is the temperature at
a reference radius $r_*$ (we take below $r_*=4\;r_h$). This kind of behavior
is expected in the case of a geometrically
thin disk but is certainly a very rough approximation in the case of a 
thick disk. \\
\subsection{Collapsars :}
In collapsars
the accretion flow toward the black hole 
has been studied in detail by \cite{popham:99}. 
The inner disk is fed by material from the collapsing stellar enveloppe.
It is less dense than in the merger case (typical densities are $10^{8}\ \mathrm{g/cm^{3}}$) 
 and is
 optically thin \citep{macfadyen:99}. 
We follow \citet{popham:99} to obtain 
an analytical expression for the 
temperature 
from the balance between the dissipated energy and neutrino losses 
\begin{equation}
\dot{q}_{eN}=Q T^6 =\frac{9}{4}\nu\;\Omega_\mathrm{K}^2
\end{equation}
where $\dot{q}_{eN}$ is the cooling rate per unit mass due to the emission of 
neutrinos by nucleons ($Q=1.4\,10^{18}$ with $T$ in MeV); 
$\dot{q}_{eN}$ is the dominant cooling contribution
as long as the disk temperature does not exceed about 10 MeV. 
We adopt an $\alpha$-prescription for the disk viscosity
\begin{equation}
\nu=\alpha H v_\mathrm{s}
\end{equation}
where $H=\frac{v_\mathrm{s}}{\Omega_\mathrm{K}}$ is the disk half 
thickness and $v_\mathrm{s}$ is
the sound velocity. If the perfect gas contribution dominates in the disk
then, $v_\mathrm{s}^2\sim \frac{\mathcal{R}T}{\mu}$ ($\mu$ being the average molecular 
weight) and the disk temperature is given by
\begin{equation}
T_d(r)\simeq 2\,\mu_\mathrm{BH}^{-0.2}\left(\frac{\alpha}{0.01}\right)^{0.2}
\left(\frac{r}{r_*}\right)^{-0.3}\ \ \ \mathrm{MeV}\ .
\label{eq:TThin}
\end{equation}
\section{Dynamics of the wind from the disk to the sonic point}
\label{sec:Dynamics}
\subsection{Wind equations}
\begin{figure}
\vspace*{-12ex}

\resizebox{\hsize}{!}{\includegraphics{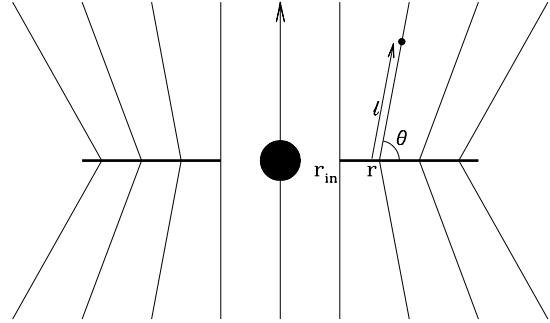}}
\vspace*{-16ex}

\caption{Schematic view of the system geometry. The disk is assumed to 
be geometrically thin and the magnetic field lines make an angle 
$\theta$ with the disk. The outflow follow the field lines and the
position of a fluid element along a line anchored at a radial distance $r$
from the black hole is represented by $y=\ell/r$.}
\label{fig:SchematicView}
\end{figure}
To study the dynamics of the wind we make a number of simplifying assumptions.
We first suppose that the inner disk is geometrically thin 
which is certainly wrong in the merger case and probably a poor approximation in  the 
collapsar scenario.
Wind material is
guided along magnetic field lines for which we adopt the simplest
possible geometry:
close to the disk the field is poloidal,
made of straight lines making an 
angle $\theta(r)$ with the plane of the disk ($r$ being the distance from 
the foot of the line to the disk axis; see Fig.~\ref{fig:SchematicView}). Since we limit our study to the part of 
the flow below the sonic point (which is sufficient to obtain the mass loss 
rate) we use non relativistic equations ($v/c\ll1$) but we adopt a
Paczynski-Wiita potential for the black hole. We also
assume that the wind has reached a stationary
regime.
Clearly, a realistic description would imply a thick disc, a complicated field geometry and a time-dependent wind  dynamics 
but we believe that the toy model presented in this paper remains able to identify the main physical processes which affects the baryonic pollution of the wind.\\

We write the three flow equations in a frame co\-ro\-ta\-ting 
with 
the foot of the line:
\begin{itemize}
\item Conservation of mass :
\begin{equation}
\rho v s(y)=\dot{m}\ ,
\label{eq:Flow1}
\end{equation}
\item Euler equation :
\begin{equation}
v\frac{dv}{dy}=\gamma(y)r-\frac{1}{\rho}\frac{dP}{dy}\ ,
\label{eq:Flow2}
\end{equation}
\item Energy equation :
\begin{equation}
v\frac{de}{dy}=\dot{q}(y)r+v\frac{P}{\rho^2}
\frac{d\rho}{dy}\ ,
\label{eq:Flow3}
\end{equation}
\end{itemize}
where $y=\ell/r$, $\ell$ being the distance along the field line 
($\ell=0$ in the plane of the disk); $e$ is the specific internal energy, 
$\gamma(y)$ and $\dot{q}(y)$ are 
respectively the total acceleration (gravitational + centrifugal) and the 
power deposited per unit mass in wind material due to neutrino heating
(and cooling), 
viscous or ohmic dissipation and
magnetic reconnection. Because the field and
stream lines are coincident the function $s(y)$ can be simply related to 
the field geometry. We 
get 
\begin{equation}
s(y)=1+a y+b y^2
\label{eq:Surf}
\end{equation}
with
\begin{equation}
a=\cos{\theta}-\sin{\theta} \frac{d\theta}{d\log{r}}
\end{equation}
and
\begin{equation}
b=-\cos{\theta}\;\sin{\theta} \frac{d\theta}{d\log{r}}\ .
\end{equation} 
Finally, $\dot{m}$ is the mass loss rate per unit surface of the disk.
The acceleration $\gamma(y)$ is derived from the potential
\begin{equation}
\gamma(y)r = -\frac{d\Phi}{dy}\ \ \ \ \ ,\ \ \ \ \ 
\Phi(y)=\Phi_\mathrm{BH}(y)+\Phi_\mathrm{C}(y)\ ,
\label{eq:Potential}
\end{equation}
where 
\begin{eqnarray}
\Phi_\mathrm{BH}(y) & = & -\frac{GM_\mathrm{BH}}{r\sqrt{y^2+2 y \cos{\theta} +1}-r_h}\nonumber\\
& = & -\frac{c^2}{2}\left[\frac{1}{x\sqrt{y^2+2 y \cos{\theta} +1}-1}\right]
\end{eqnarray}
is the black hole potential ($x=r/r_h$) and
\begin{eqnarray}
\Phi_\mathrm{C}(y) & = & -\frac{1}{2}\Omega^2(r)r^2(1+y\cos{\theta})^2\nonumber\\
& = & -\frac{c^2}{4}\left[\frac{x}{(x-1)^2}(1+y\cos{\theta})^2\right]
\end{eqnarray}
is the centrifugal potential. There is a critical angle $\theta_\mathrm{cr}$
below which the acceleration is always positive so that matter (even at zero
temperature) can escape
freely from the disk without being confined in a potential well. In newtonian
gravity $\theta_\mathrm{cr}=60^{\circ}$ \citep{blandford:82}
but $\theta_\mathrm{cr}$ slowly 
decreases when the Paczynski-Wiita potential is used,
from about 63$^{\circ}$ at $x=3$ to 60$^{\circ}$ at large radial
distances. 
\subsection{Equation of state}
Our equation of state includes nucleons, relativistic electrons and positrons, 
and photons. Following \citet{bethe:80} and \citet{bethe:93} the contribution of 
relativistic particles is given by
\begin{eqnarray}
P_r & = & \frac{(kT)^4}{12(\hbar c)^3}\left(\frac{11 \pi^2}{15}+2\eta^2+\frac{\eta^4}{\pi^2}\right)\nonumber\\
& = & (1.26+0.35\eta^2+0.017\eta^4)10^{26}\ T^4_\mathrm{MeV}\ \mathrm{dyne.cm}^{-2}\nonumber\\
\label{eq:Pr}
\end{eqnarray}
where $\eta=\mu_e/kT$, $\mu_e$ being the electron chemical potential. Nucleons
behave as an ideal gas of pressure
\begin{equation}
P_N=\frac{\rho}{m_N}kT
\end{equation}
where $m_N$ is the nucleon mass and the density is obtained from 
\begin{eqnarray}
\rho & = & \frac{m_N}{3}{\left(\frac{kT}{\hbar c}\right)^3}Y_e^{-1}
\left(\eta+\frac{\eta^3}{\pi^2}\right)\nonumber\\
& = & 7.2\;10^7\; 
T^3_\mathrm{MeV}Y_e^{-1}\left(\eta+\frac{\eta^3}{\pi^2}\right)\
\mathrm{g.cm}^{-3}\ .
\end{eqnarray}
The number of electrons per nucleon $Y_e$ should be 
computed from the rates of neutrino capture and emission  
by nucleons. In practice, we do not perform this calculation 
and simply adopt a constant $Y_e=0.5$.
Finally, the specific internal energy is 
\begin{equation}
e=\frac{(3 P_r+3/2 P_N)}{\rho}
\end{equation}
so that the three thermodynamic quantities $P$, $\rho$ and $e$ that
appear in the flow equations can be expressed in terms of $T$ and $\eta$.
\subsection{The sonic point}
To solve the flow Eqs (\ref{eq:Flow1}--\ref{eq:Flow3}) we first derive Eq.~(\ref{eq:Flow1}) and express 
the thermodynamic variables $\rho$, $P$ and $e$ as functions of $T$ and 
$\eta$ to obtain a linear system for $log\;v$, $log\;T$ and $log\;\eta$
\begin{eqnarray}
& & \!\!\!\!\!\frac{d\log{v}}{dy} +
3\frac{d\log{T}}{dy}+A(\eta)\frac{d\log{\eta}}{dy} =
-\frac{d\log{s(y)}}{dy}\ ,\nonumber\\
& & \!\!\!\!\!v^{2}\frac{d\log{v}}{dy} +
4\frac{P}{\rho}\frac{d\log{T}}{dy}+B(\eta)\frac{d\log{\eta}}{dy} =
\gamma(y)r\ ,\nonumber\\
& & \!\!\!\!\!v\!\left[e-3\frac{P}{\rho}\right]\!\!\frac{d\log{T}}{dy} +
v\!\left[C(\eta)e-A(\eta)\frac{P}{\rho}\right]\!\!\frac{d\log{\eta}}{dy} =
\dot{q}(y) r\ ,\nonumber\\
\end{eqnarray}
where we have used the equation of state to get
\begin{eqnarray}
\left.\frac{\partial\log{\rho}}{\partial\log{T}}\right|_{\eta} = 3 & ,
& \left.\frac{\partial\log{\rho}}{\partial\log{\eta}}\right|_{T} =
A(\eta)\ \ ,\nonumber\\
\left.\frac{\partial\log{P}}{\partial\log{T}}\right|_{\eta} = 4 & , &
\left.\frac{\partial\log{P}}{\partial\log{\eta}}\right|_{T} = B(\eta)\
\ ,\nonumber\\
\left.\frac{\partial\log{e}}{\partial\log{T}}\right|_{\eta} = 1 & , &
\left.\frac{\partial\log{e}}{\partial\log{\eta}}\right|_{T} = C(\eta)\ \ .
\end{eqnarray}
The three derivatives of $v$, $T$ and $\eta$ can then be written as
\begin{eqnarray}
\frac{d\log{v}}{dy}    & = & \frac{F_1(y,v,T,\eta)}{\Delta}\ ,\nonumber\\
\frac{d\log{T}}{dy}    & = & \frac{F_2(y,v,T,\eta)}{\Delta}\ ,\nonumber\\
\frac{d\log{\eta}}{dy} & = & \frac{F_3(y,v,T,\eta)}{\Delta}\ ,
\label{eq:Derivatives}
\end{eqnarray}
where $\Delta(v,T,\eta)$ is the determinant of the system. 
It is equal to zero at the sonic point which gives
\begin{equation}
v^2=v_\mathrm{s}^2=\left[\frac{(B-4C)-(\tilde{\gamma}-1)(3B-4A)}{(A-3C)}\right]\frac{P}{\rho}
\label{eq:Sonic1}
\end{equation}
where $\tilde{\gamma}-1={P}/{\rho\,e}$. If the equation of state is dominated 
by the contribution of relativistic (resp. non relativistic) particles
$\tilde{\gamma}-1=1/3$ (resp. 2/3) and 
$v_\mathrm{s}^2=\frac{4}{3}(\mathrm{resp.}\ \frac{5}{3})\frac{P}{\rho}$.\\

Since the three
functions $v$, $T$ and $\eta$ must remain re\-gu\-lar everywhere in the wind the
numerators in Eq.~(\ref{eq:Derivatives}) must be zero at the sonic point
\begin{eqnarray}
F_1(y_\mathrm{s},v_\mathrm{s},T_\mathrm{s},\eta_\mathrm{s})\!=\!
F_2(y_\mathrm{s},v_\mathrm{s},T_\mathrm{s},\eta_\mathrm{s})\!=\!
F_3(y_\mathrm{s},v_\mathrm{s},T_\mathrm{s},\eta_\mathrm{s})\!=\!0\nonumber\\
\end{eqnarray}
which yields a unique relation among the parameters at the sonic point
\begin{eqnarray}
v_\mathrm{s}^2\left.\frac{d\log{s(y)}}{dy}\right|_{y_\mathrm{s}}\!\!\!+\gamma(y_\mathrm{s})r
-\frac{\dot{q}(y_\mathrm{s}) r}{v_\mathrm{s}}\left(\frac{P}{\rho\,e}\;
\frac{4A-3B}{A-3C}\right)_{T_\mathrm{s},\eta_\mathrm{s}}\!\!\!\!\!\!=0\nonumber\\
\label{eq:Sonic2}
\end{eqnarray}
the parenthesis being equal to 1/3 (resp. 2/3) for relativistic (resp. non
relativistic) particles. Since ({\it i}) the last term 
in Eq.~(\ref{eq:Sonic2}) is smaller 
than the two others in most cases of interest and ({\it ii}) 
the derivative of $s(y)$ is positive (if the inclination angle 
$\theta(r)$ of the field lines decreases with increasing distance to the 
axis) it can be seen that the sonic point is 
located in the region where $\gamma(y)$ is negative, i.e. below 
$y_1$ where $\gamma(y_1)=0$. It appears in practice that $y_\mathrm{s}$ is 
very close to $y_1$ (except na\-tu\-ral\-ly for $\theta=90^{\circ}$ for which
$y_1\rightarrow \infty$). The difference is typically less than 1\% even
for $\theta=89^{\circ}$.
\subsection{Heating and cooling sources}
Several sources can contribute to the injection of energy in wind material: 
viscous or ohmic dissipation, magnetic reconnection or neutrino 
processes (capture on free nucleons, scattering on electrons and positrons 
or neutrino-antineutrino annihilation). Cooling occurs through neutrino emission
by nucleons and annihilation of electron-positron pairs. 
A detailed description of all these processes is beyond the scope 
of this paper and we have rather considered two limiting cases in a very
simplified way.\\
When the disk is optically thin to neutrinos we adopt a uniform heating 
(per unit mass)
$\dot{q}_{h}$ along the field line and
limit the cooling to neutrino emission by nucleons i.e.
\begin{equation}
\dot{q}=\dot{q}_{h}-QT^6
\label{eq:HeatingThin}
\end{equation}
Although this is not strictly satisfied in realistic models \citep{popham:99,macfadyen:99}, we require that $\dot{q}=0$ in the plane of the disk
which fixes $\dot{q}_{h}$
\begin{equation}
\dot{q}_{h}=Q T_0^6
\label{eq:Equality}
\end{equation}
where $T_0$ is the disk temperature at the foot of the line given by Eq.~(\ref{eq:TThin}). \\
In the case of an optically thick disk we make the extreme assumption that all the 
heating which is not due to neutrino processes takes place below the 
neutrinosphere. The power $\dot{q}$ injected into the wind is then restricted
to the neutrino contributions
\begin{equation}
\dot{q}=\dot{q}_{\nu}=\dot{q}_{\nu N}+\dot{q}_{\nu e}
+\dot{q}_{\nu \bar{\nu}}-\dot{q}_{e N}-\dot{q}_{e^+e^-}
\label{eq:Heating}
\end{equation}
where the different terms in Eq.~(\ref{eq:Heating}) respectively correspond to  
neutrino capture on free nucleons, scattering on electrons and positrons, 
neutrino-antineutrino-annihilation (heating) neutrino emission
by nucleons and annihilation of electron-positron pairs (cooling).
With our assumption that 
$Y_e=Y_p=Y_n=0.5$ the electron neutrino and antineutrino temperatures are 
identical and the 
heating by capture on free nucleons takes the simple form \citep{qian:96}
\begin{eqnarray}
\dot{q}_{\nu N} & = & Q\int_\mathrm{disk}\left[I_{\nu_e}\frac{\langle\epsilon_{\nu_e}^3\rangle}{\langle\epsilon_{\nu_e}\rangle}
+I_{\bar{\nu}_e}\frac{\langle\epsilon_{\bar{\nu}_e}^3\rangle}{\langle\epsilon_{\bar{\nu}_e}\rangle}\right]\;
d\Omega\nonumber\\
& = & 1.4\;10^{18}\int_\mathrm{disk}T_{\nu}^6\;\frac{d\Omega}{4\pi}\ \ \ \mathrm{erg.g^{-1}.s^{-1}}
\label{eq:HeatingNuN}
\end{eqnarray}
where the temperature is in MeV, $I_{\nu_e}=I_{\bar{\nu}_e}=\frac{7}{16}
\frac{\sigma T_{\nu}^4}{\pi}$ is the neutrino (antineutrino) intensity 
and $\langle\epsilon_{\nu_e}^n\rangle=\langle\epsilon_{\bar{\nu}_e}^n\rangle$ 
are the $n$th energy moments of the neutrino Fermi distribution.
The integral is performed over the
disk surface and $d\Omega$ is the solid angle of a surface element as seen from
a point of coordinate $y$ on the field line. All relativis\-tic effects 
on the neutrinos (bending of trajectories, gravitational and
Doppler shifts) have been neglected even if they can lead to appreciable
corrections in the results \citep{jaroszynski:93}. 
The cooling by the reverse 
reactions (neutrino emission by nucleons) is given by
\begin{equation}
\dot{q}_{eN}=1.4\;10^{18}\;T^6\ \ \ \mathrm{erg.g^{-1}.s^{-1}}
\end{equation}
where $T$ is the local temperature in the wind. Pauli blocking effects for
electrons have been neglected since $\eta\sim 0.1$ everywhere in our wind
solutions except in the vicinity of the disk. To compute the heating
rate due to neutrino scattering on relativistic electrons and positrons
we use the expression given
by \citet{herant:94} adapted to the disk geometry
\begin{equation}
\dot{q}_{\nu e}=3.6\;10^{24} \frac{T^4}{\rho}\int\limits_\mathrm{disk}
(T_{\nu}-T)T_{\nu}^4\;d\Omega\ \ \ \mathrm{erg.g^{-1}.s^{-1}}
\end{equation}
The heating rate by neutrino-antineutrino annihilation is obtained from
a double integral over the disk surface
\begin{eqnarray}
\!\dot{q}_{\nu\bar{\nu}} & = & \!\frac{1}{\rho}\left\lbrace\!Q_1\!\!{\int\limits_\mathrm{disk}\!\!d\Omega\;I_{\nu_e}}\!\!{\int\limits_\mathrm{disk}\!\!d\Omega^{\prime}\;I_{\bar{\nu}_e}^{\prime}}\!\left[\frac{\langle\epsilon_{\nu_e}^2\rangle}{\langle\epsilon_{\nu_e}\rangle}+\frac{\langle\epsilon_{\bar{\nu}_e}^2\rangle^{\prime}}{\langle\epsilon_{\bar{\nu}_e}\rangle^{\prime}}\right]\!(1\!-\!\cos{\alpha})^2\right.\nonumber\\
& & \left.+Q_2\!{\int\limits_\mathrm{disk}\!\!d\Omega\;I_{\nu_e}}\!\!{\int\limits_\mathrm{disk}\!\!d\Omega^{\prime}\;I_{\bar{\nu}_e}^{\prime}}
\frac{\langle\epsilon_{\nu_e}\rangle+\langle\epsilon_{\bar{\nu}_e}\rangle^{\prime}}{\langle\epsilon_{\nu_e}\rangle \langle\epsilon_{\bar{\nu}_e}\rangle^{\prime}}\;(1\!-\!\cos{\alpha})\right\rbrace\nonumber\\
\label{eq:HeatingNuNuBar}
\end{eqnarray}
where $Q_1$ and $Q_2$ are two constants given in \citet{ruffert:97}. 
The prime quantities correspond to a surface \mbox{element}
$dS^{\prime}$ whose neutrinos interact with those emitted by a surface element
$dS$, $\alpha$ being the interaction angle.
When 
the neutrino intensities and average energies are \mbox{expressed} as a function 
of the neutrinosphere temperature $T_{\nu}$ Eq.~(\ref{eq:HeatingNuNuBar}) becomes
\begin{eqnarray}
\dot{q}_{\nu\bar{\nu}} & = & \frac{1}{\rho} \left\lbrace 1.6\
10^{22}\!\!\int\limits_\mathrm{disk}\!\!d\Omega\;T_{\nu}^4\!\!\int\limits_\mathrm{disk}\!\!d\Omega^{\prime}\;{T_{\nu}^{\prime}}^4
(T_{\nu}+T_{\nu}^{\prime}) (1\!-\!\cos{\alpha})^2\right.\nonumber\\
& & \left.+\;6.7\ 10^{20}\!\!\int\limits_\mathrm{disk}\!\!d\Omega\;T_{\nu}^4\!\!\int\limits_\mathrm{disk}\!\!d\Omega^{\prime}\;{T_{\nu}^{\prime}}^4
\frac{T_{\nu}+T_{\nu}^{\prime}}{T_{\nu}T_{\nu}^{\prime}}
(1\!-\!\cos{\alpha})\right\rbrace\nonumber\\
& & \mathrm{erg.g^{-1}.s^{-1}}
\end{eqnarray}
Finally, the cooling rate from the annihilation of electron-positron pairs 
is given by \citep{herant:94}
\begin{equation}
q_{e^+e^-}=1.5\;10^{25}\frac{T^9}{\rho}\ \ \ \mathrm{erg.g^{-1}.s^{-1}}
\label{eq:HeatingEPEM}
\end{equation}
\section{The mass loss rate}
\label{sec:MDot}
\subsection{Numerical solution for an optically thin disk}
\label{sec:NumericalThin}
\begin{figure}
\resizebox{\hsize}{!}{\includegraphics{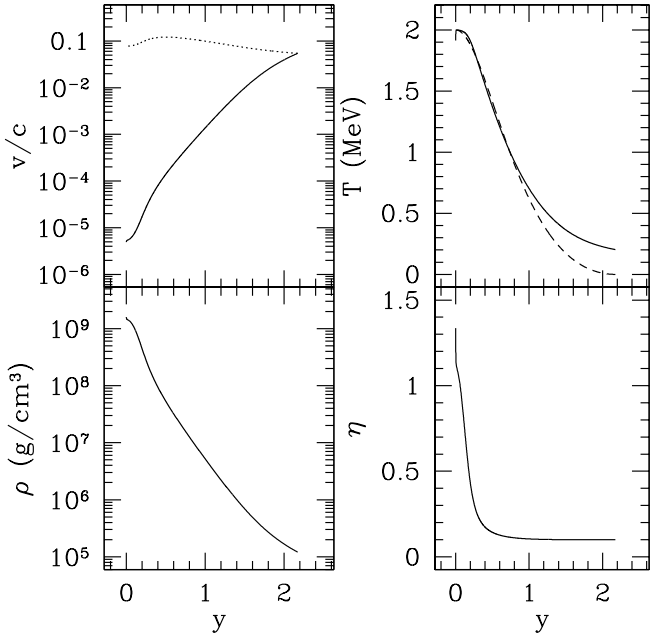}}
\caption{Wind solution for an optically thin disk with uniform heating. 
The field line is anchored at $r_*=4\,r_h$ 
and makes an angle $\theta=85^{\circ}$ with the disk. The temperature
at the foot of the line is $T_d=2$ MeV.
In the velocity plot the dotted line shows the local sound velocity. The 
sonic point is located at $y_s=2.174$. The dashed line in the temperature 
plot corresponds to the approximate analytical solution (Eq.~(\ref{eq:TApp})). 
The electron
degeneracy parameter $\eta=\mu_e/kT$ is close to 0.1 except at the 
vicinity of the disk. }
\label{fig:SolutionThin}
\end{figure}
We first solve the flow equations for an optically thin disk and in the case 
of uniform heating, i.e. with $\dot{q}$ given by Eq.~(\ref{eq:HeatingThin}).
We obtain the mass 
loss rate $\dot{m}$ in a
classical way by integrating inward from the sonic point down to the disk 
surface. We fix trial values of $T_\mathrm{s}$ and $\eta_\mathrm{s}$ at the sonic 
point from which we get $v=v_\mathrm{s}$ from Eq.~(\ref{eq:Sonic1}) and the position $y_\mathrm{s}$ 
from Eq.~(\ref{eq:Sonic2}). To start the integration we need the derivatives of $v$,
$T$ and $\eta$ at $y=y_\mathrm{s}$. These derivatives cannot be directly 
calculated from Eq.~(\ref{eq:Derivatives}) because $F_1=F_2=F_3=\Delta=0$ at the sonic point. We instead
use l'H\^opital's rule which allows us to write three algebraic, 
second-order equations for the three derivatives and we solve these equations
with a Newton-Raphson technique. Once $v$, $T$, $\eta$ and their 
derivatives have been determined at the sonic point the inward integration can 
be started. In agreement with the results of \citet{duncan:86} for 
neutrino-driven winds in proto-neutron stars we observe that
at some position $y=y_*$ the velocity begins to fall off rapidly while the 
temperature reaches a maximum $T_\mathrm{max}\le T_d(r)$ 
($T_d(r)$ being the disk temperature at radius $r$). 
We then adjust the 
values of $T_\mathrm{s}$ and $\eta_\mathrm{s}$ with the requirement that 
$y_*$ should be as close as possible to 0 and $T_\mathrm{max}$ to $T_d(r)$.\\

We first constructed a reference model where we follow the wind along a
field line attached at $r=r_*=4\;r_h$. The disk temperature at $r=r_*$ is 
$T_d=2$ MeV. The line makes an angle $\theta=85^{\circ}$ with the disk
and the derivative $\frac{d\theta}{dr}=0$. The mass of the black hole 
is $M_\mathrm{BH}=2.5\ \mathrm{M}_{\sun}$.
The results for this
reference model are shown 
in Fig.~\ref{fig:SolutionThin}.
The sonic point is located at $y_\mathrm{s}=2.174$ slightly below
$y_1=2.182$ where $\gamma=0$. The temperature and density at the sonic point 
are
$T_\mathrm{s}=0.203$ MeV and $\rho_\mathrm{s}=1.2\;10^5$ g.cm$^{-3}$ which 
correspond to
a sound velocity $v_\mathrm{s}/c=5.44\,10^{-2}$ or $v_\mathrm{s}=16300$ km.s$^{-1}$
and a mass loss rate  
$\dot{m}=2.3\;10^{14}$ g.cm$^{-2}$.s$^{-1}$. The degeneracy parameter $\eta$
remains 
practically constant ($\eta\simeq 0.1$) from $y=0.5$ to the sonic point, a property 
which will be 
used to construct the analytical solutions in Sect.~\ref{sec:Analytical} below. Neutrino 
cooling is efficient close to the disk, up to $y\la 0.5$.
We have tested the effect of a different field line geometry with 
$\theta^{\prime}=\frac{d\theta}{dr}\ne 0$. We considered 
two cases,
$\theta^{\prime}=-2^{\circ}/r_h$ and $-5^{\circ}/r_h$, for which we respectively
obtain $\dot{m}=2.5$ and $2.8\;10^{14}$ g.cm$^{-2}$.s$^{-1}$.\\

\begin{figure*}
\begin{tabular}{cc}
\resizebox{0.45\hsize}{!}{\includegraphics{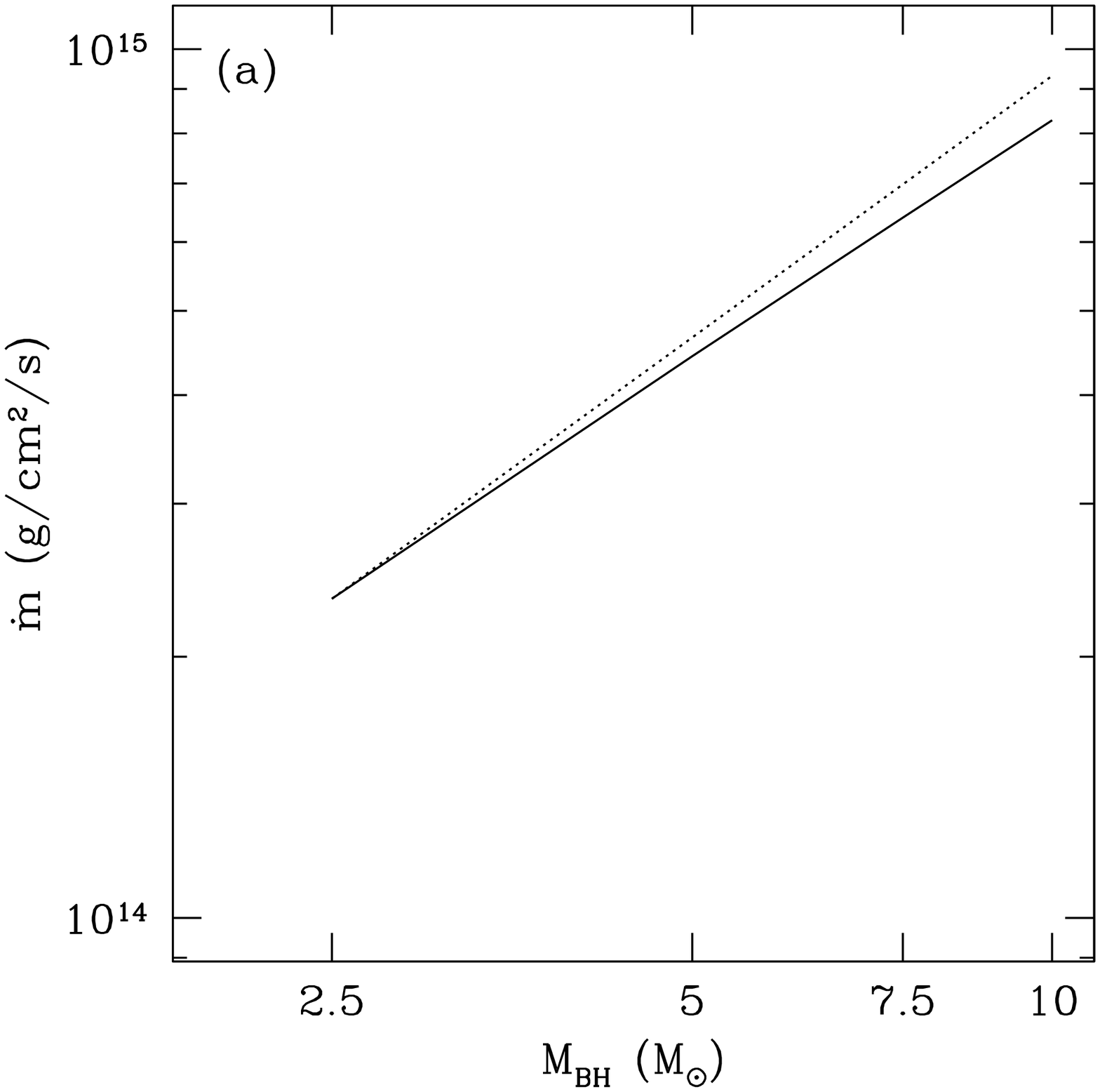}} & \resizebox{0.45\hsize}{!}{\includegraphics{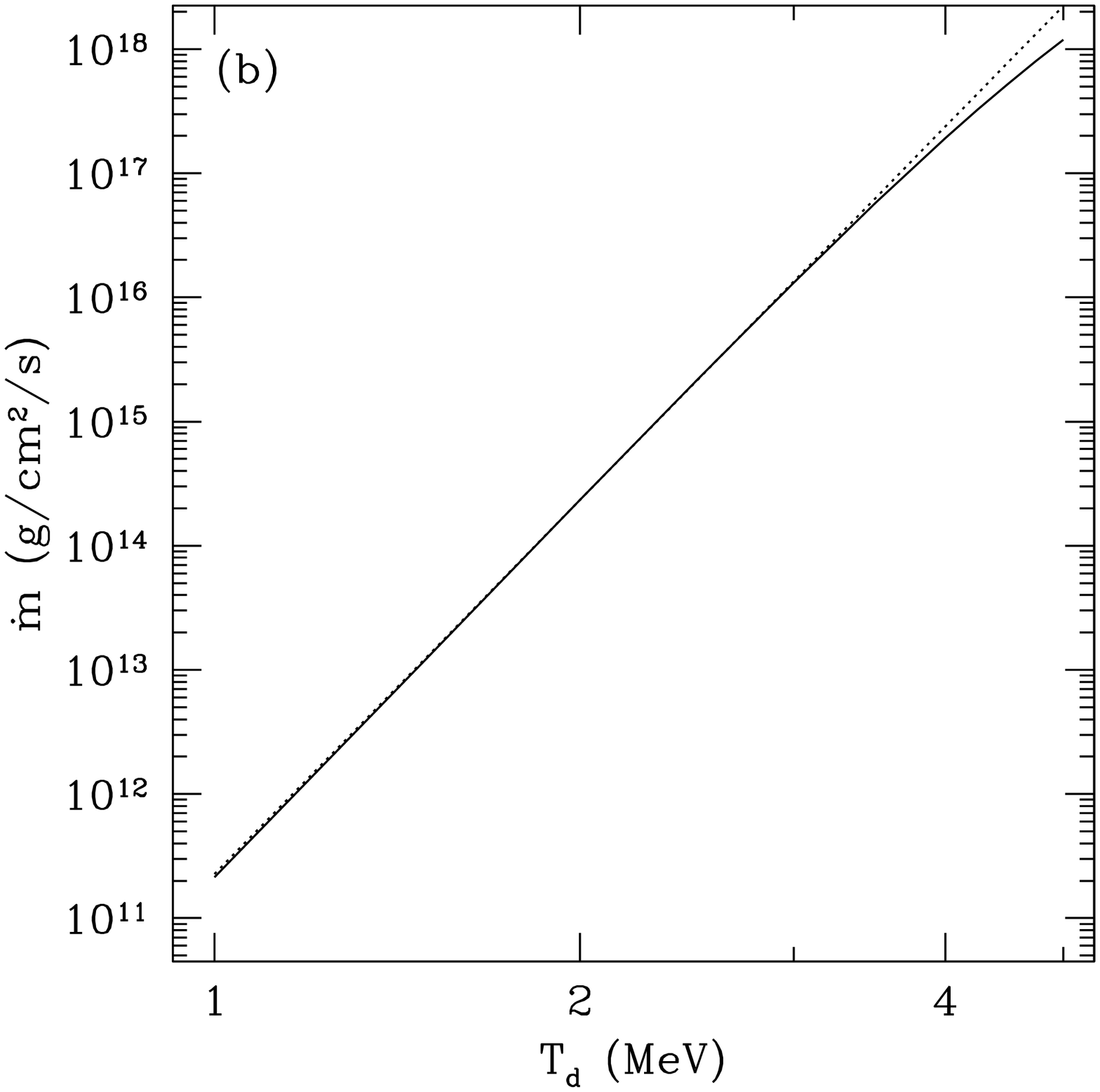}}\\
\resizebox{0.45\hsize}{!}{\includegraphics{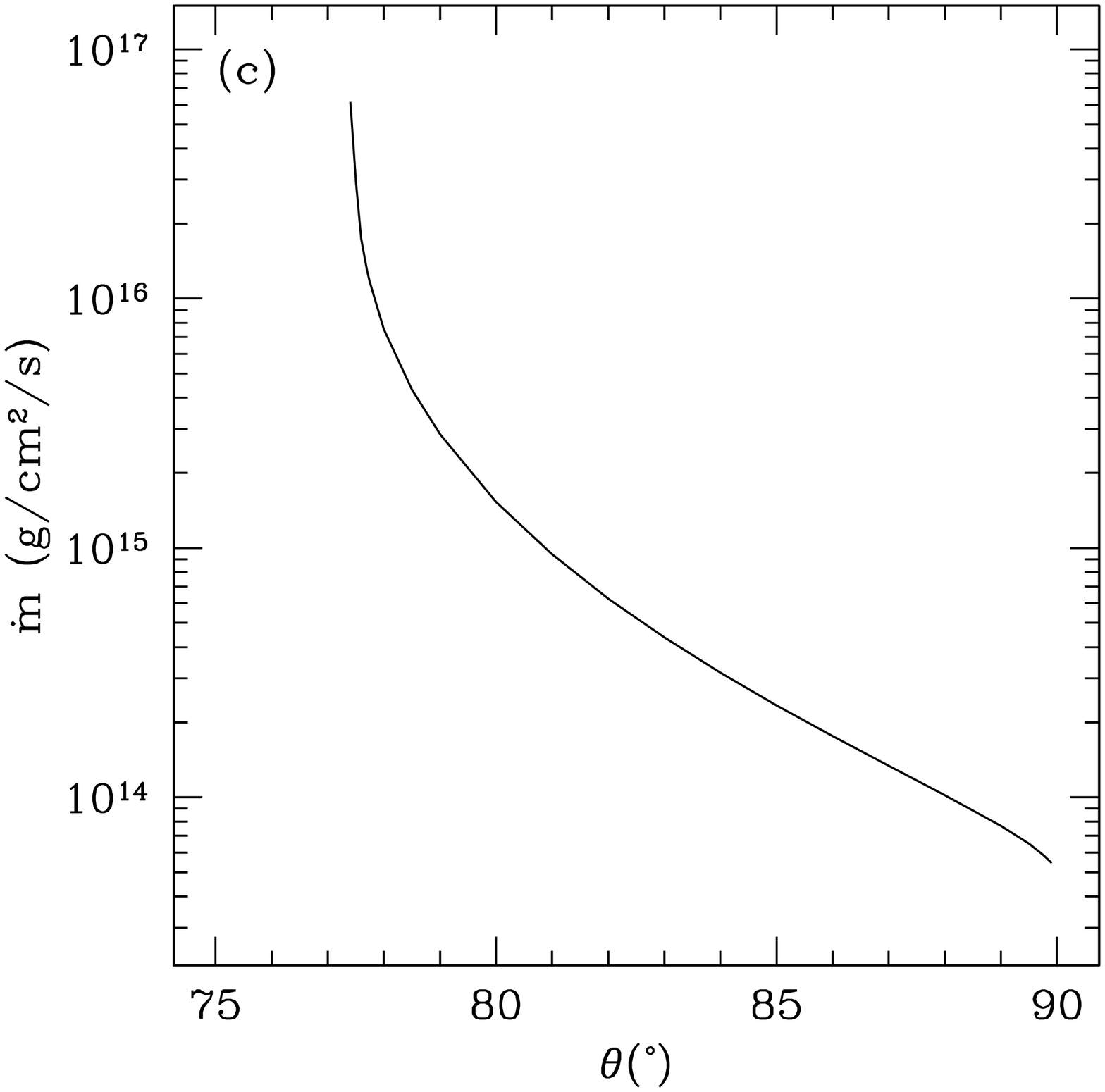}} & \resizebox{0.45\hsize}{!}{\includegraphics{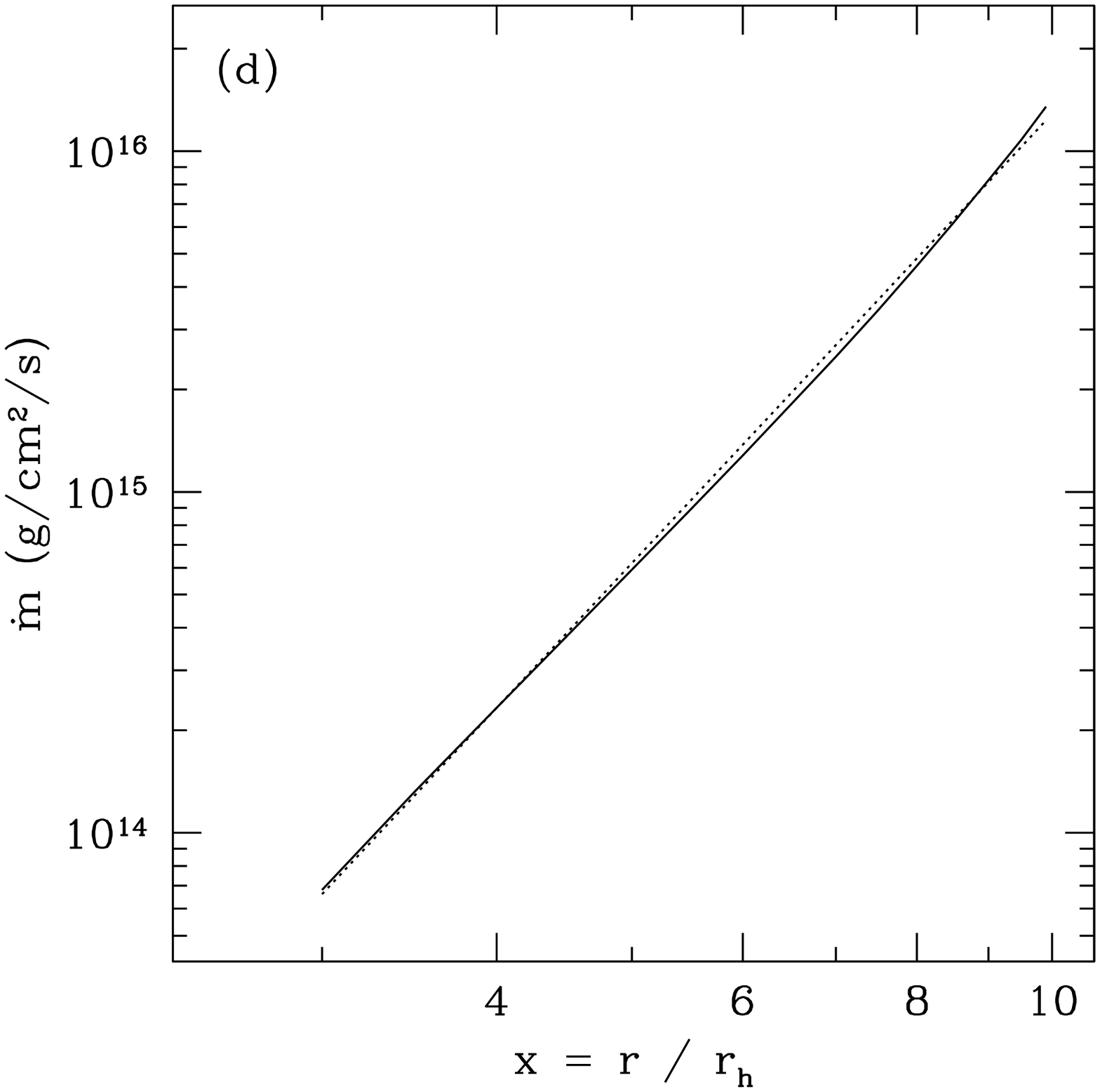}}
\end{tabular}
\caption{Mass loss rate per unit surface 
for an optically thin disk with uniform heating as a function
of (a) black hole mass $M_\mathrm{BH}$, (b) disk temperature at the foot of the
line $T_d$, (c) inclination angle $\theta$ and (d) 
radial distance to the black hole. When one quantity
is varied the three others are maintained at their reference values:
$M_\mathrm{BH}=2.5\ \mathrm{M}_{\sun}$, $T_d=2$ MeV, $\theta=85^{\circ}$ and $r=4\,r_h$.
The dotted lines in (a), (b) and (d) have respective slopes 1, 10 and 4.4.}
\label{fig:Dependance}
\end{figure*}

\begin{figure}
\resizebox{\hsize}{!}{\includegraphics{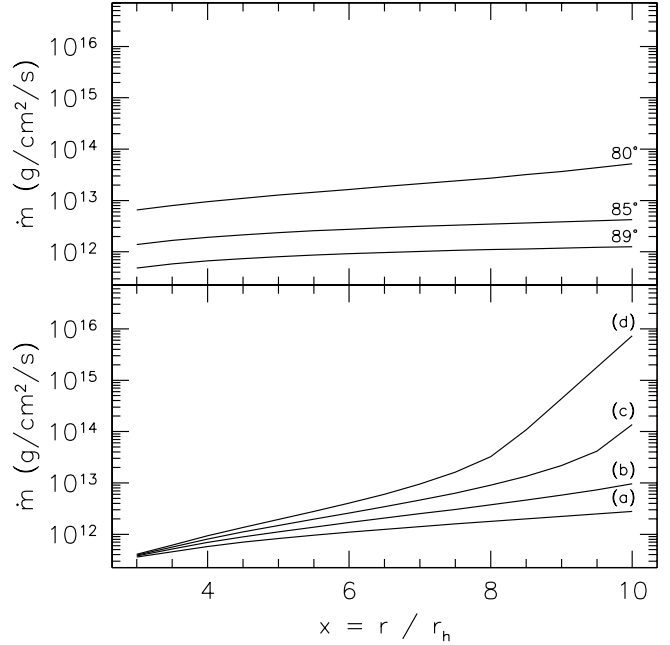}}
\caption{Mass loss rate per unit surface as a function of
$x=r/r_h$ for an optically thin disk with uniform heating.
Upper panel: field lines with constant inclination angles $\theta=80$, 85 and 
89$^{\circ}$; lower panel: field lines with decreasing inclination angle
$\theta=90^{\circ}-\lambda(x-3)$ with $\lambda=0.5$, 1, 1.5 and 2$^{\circ }$ (curves 
labelled (a) to (d)).}
\label{fig:MDotThin}
\end{figure}

We then varied the mass of the black hole $M_\mathrm{BH}$, the disk
temperature $T_d$, the inclination angle $\theta$ and the position $r$ of the foot of the 
line 
to see how these parameters 
affect the mass loss rate. The results are shown in
Fig.~\ref{fig:Dependance} where we have plotted $\dot{m}$ when 
$M_\mathrm{BH}$, $T_d$, $\theta$ and $x=r/r_h$ are varied 
separately while the other three quantities are maintained at fixed values
(choosen to be those of the re\-fe\-rence model: 
$M_\mathrm{BH}=2.5\ \mathrm{M}_{\sun}$, $T_d=2$ MeV, $\theta=85^{\circ}$ and $x=4$). 
The mass loss rate appears to be nearly proportional to
the mass of the black hole (Fig.~\ref{fig:Dependance}a). 
The dependence of $\dot{m}$ on the temperature $T_d$ at the foot of 
the field line is much more spectacular since we get 
$\dot{m}\propto T_d^{10}$ (Fig.~\ref{fig:Dependance}b) in agreement with the results for 
neutrino-driven winds in neutron stars \citep{duncan:86}. 
The mass loss rate also sharply 
increases when the angle between 
the field line and the disk is reduced (Fig.~\ref{fig:Dependance}c). 
Below $\theta\simeq 77^{\circ}$ it becomes
more and more difficult to construct wind solutions with the required 
accuracy. This can be related to the value of the Bernouilli function
\begin{equation}
\mathcal{B}=\frac{1}{2} v^2 + h + \Phi - \Phi_1 
\end{equation}
($h$ being the specific enthalpy and $\Phi_1$ the potential at $y=y_1$) 
which is positive in the plane of the 
disk for $\theta\le 77^{\circ}$. The initial thermal energy is then 
sufficient to allow the gas to escape even in the absence of additional heating.
Finally, the mass loss rate approximately increases as $r^{4.4}$ (Fig.~\ref{fig:Dependance}d) 
when the potential well becomes shallower at larger radial
distances.\\
As long as the Bernouilli function is not too close to zero
the mass, temperature and geometrical dependence can be separated in 
$\dot{m}$ to  
yield 
the general expression 
\begin{eqnarray}
& & \!\!\!\!\dot{m}(x)\approx 2.3\;10^{14}\mu_\mathrm{BH}\left[\frac{T_d(x)}{2 \ \mathrm{MeV}}
\right]^{10}\!f[x,\theta(x)]\ \mathrm{g.cm^{-2}.s^{-1}}\nonumber\\
\end{eqnarray}
where $f$ is a geometrical function which satisfies $f(4, 85^{\circ})=1$. 
The mass loss rate has been represented in Fig.~\ref{fig:MDotThin} as a function of $x$ 
for $\mu_\mathrm{BH}=1$, $T_d(x)$ given by Eq.~(\ref{eq:TThin}) and different field
geometries: constant inclination angles $\theta=80$, 85 and $89^{\circ}$
or $\theta$ linearly decreasing from $90^{\circ}$ at $x=3$, i.e. 
$\theta=90^{\circ}-\lambda(x-3)$~ (with $\lambda=0.5,\ 1.0,\ 1.5$ or $2^{\circ}$).
At constant $\theta$ the mass loss rate varies rather slowly with $x$ while
for $\lambda=1.5$ or $2^{\circ}$ it rapidly increases at large radial distances 
where $\theta\le 80^{\circ}$. 

\subsection{Numerical solution for an optically thick disk}
\begin{figure}
\resizebox{\hsize}{!}{\includegraphics{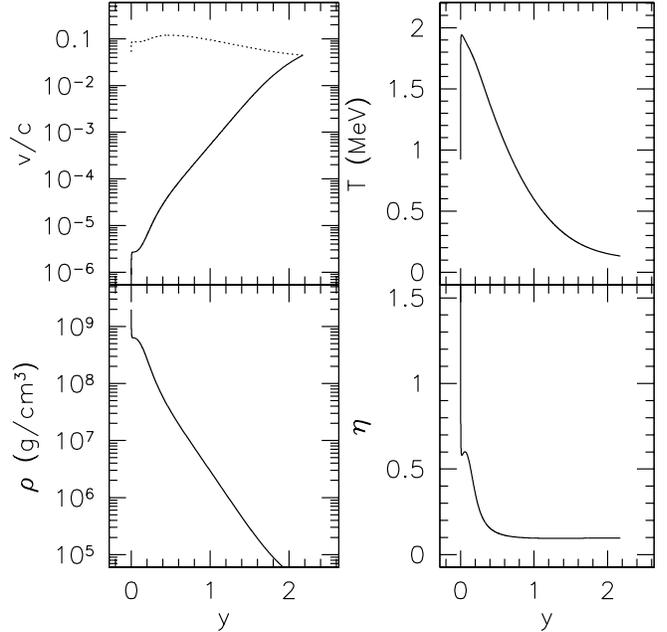}}
\caption{Same as Fig.~\ref{fig:SolutionThin} for an optically thick disk.  }
\label{fig:SolutionThick}
\end{figure}
\begin{figure}
\resizebox{\hsize}{!}{\includegraphics{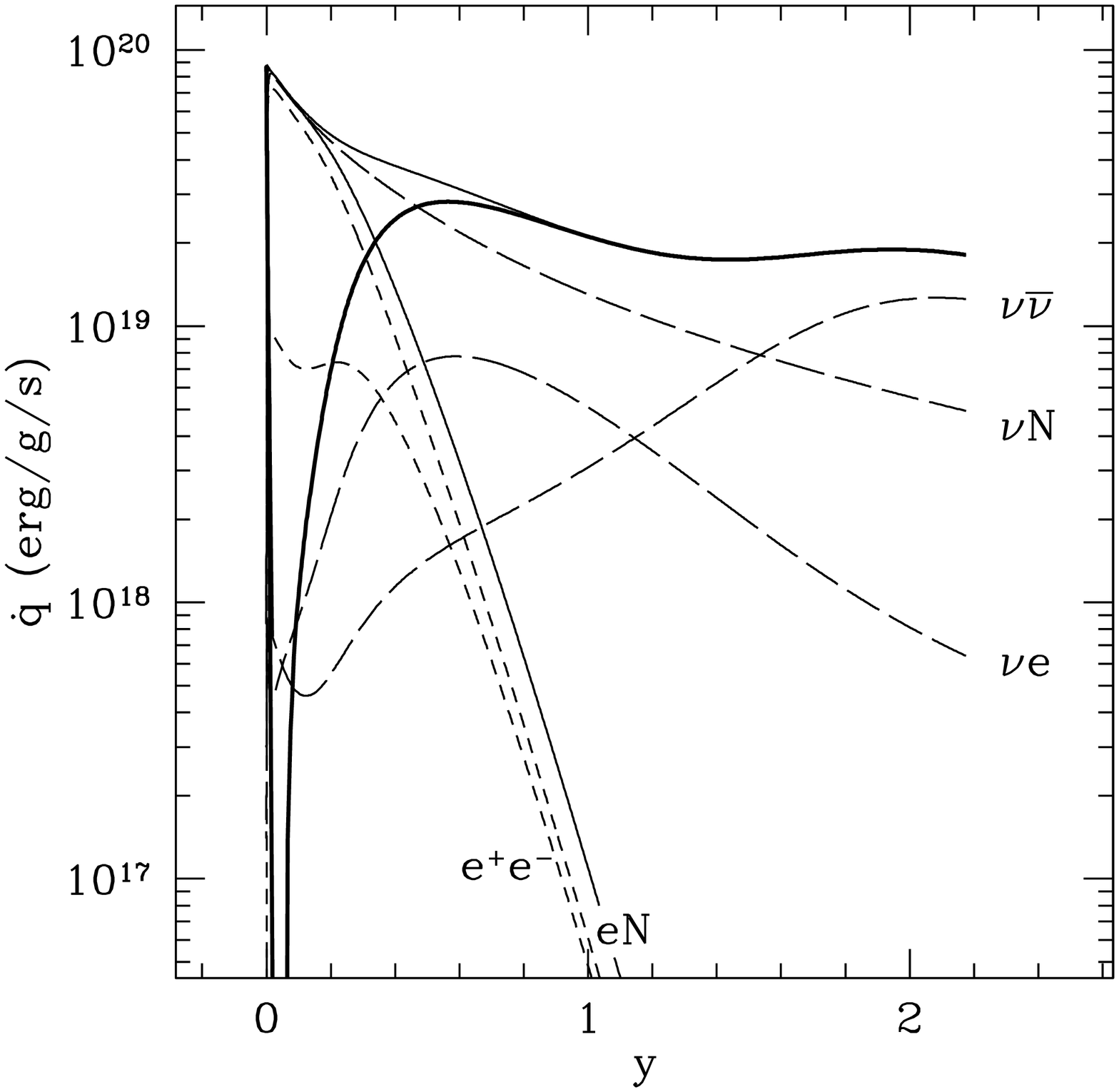}}
\caption{Neutrino heating and cooling contributions. The long and short dashed 
lines respectively
represent individual heating and cooling processes; 
$\nu N$: neutrino captures on nucleons;
$\nu e$: neutrino captures on electrons; $\nu{\bar \nu}$:
neutrino-antineutrino annihilation; $eN$: neutrino emission by nucleons;
$e^+e^-$: annihilation of electron-positron pairs.
The full thin lines represent the total of the heating ($\nu N$ + 
$\nu e$ + $\nu{\bar \nu}$) and cooling ($eN$ + $e^+e^-$) processes. 
The full thick line
is the sum of all contributions.}
\label{fig:Heating}
\end{figure}
\begin{figure}
\resizebox{\hsize}{!}{\includegraphics{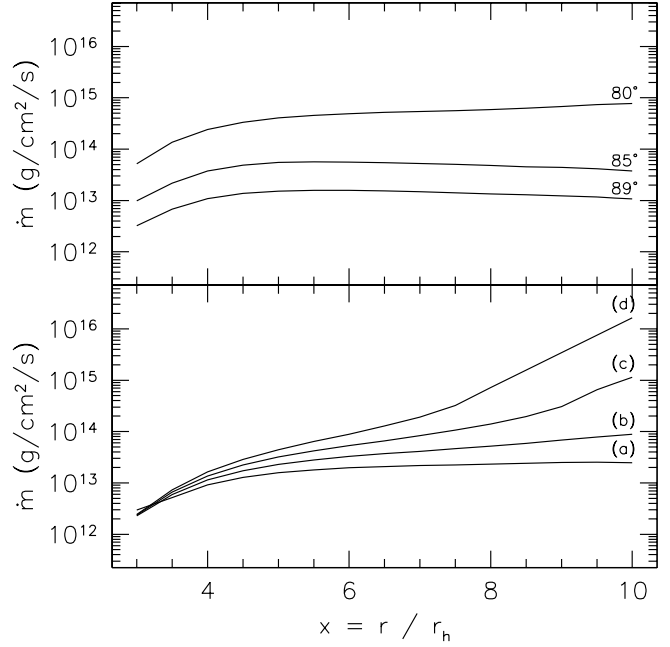}}
\caption{Same as Fig.~\ref{fig:MDotThin} for an optically thick disk.  }
\label{fig:MDotThick}
\end{figure}

If the disk is optically thick the heating and cooling sources due to 
neutrinos are specified by Eq.~(\ref{eq:HeatingNuN}--\ref{eq:HeatingEPEM}). Again, wind 
solutions are found by inward integration from the sonic point down to
the disk. 
A reference model is constructed with the same black hole mass and field 
geometry ($M_\mathrm{BH}=2.5\ \mathrm{M}_{\sun}$ and field line attached at 
$r=r_*=4\;r_h$ and making an angle $\theta=85^{\circ}$ with the disk). 
The disk is supposed to be optically thick to neutrinos from 
$r_\mathrm{in}=3\;r_h$ to $r_\mathrm{out}=10\;r_h$.
The temperature of the neutrinosphere in this reference model 
is $T_{\nu}=2$ MeV 
and does not vary with
radius which allows a simple calculation of the geometric integrals 
appearing in the neutrino heating terms. The results for this
re\-fe\-ren\-ce model are shown 
in Fig.~\ref{fig:SolutionThick}. The sonic point is located at $y_\mathrm{s}=2.175$. 
The temperature and density at the sonic point are
$T_\mathrm{s}=0.132$ MeV and $\rho_\mathrm{s}=3.23\;10^4$ g.cm$^{-3}$ which 
correspond to 
a sound velocity 
$v_\mathrm{s}/c=4.45\,10^{-2}$ or $v_\mathrm{s}=13350$ km.s$^{-1}$
and a mass loss rate  
$\dot{m}=5.1\;10^{13}$ g.cm$^{-2}$.s$^{-1}$. The neutrino 
heating and cooling terms are detailed in Fig.~\ref{fig:Heating}. The major contribution to the
heating comes from neutrino captures on nucleons while neutrino emission by 
nucleons
and annihilation of electron-positron pairs have comparable effects on the 
cooling.
We have then abandoned the assumption of constant neutrino
tem\-pe\-ra\-tu\-re which was essentially made for a rapid calculation of the 
integrals in the
neutrino heating terms. When $T_{\nu}$ depends on the radial distance in the 
disk, the calculation of $\dot{q}_{\nu{\bar \nu}}$ becomes very time consuming.
Since $\dot{q}_{\nu{\bar \nu}}$ is not the dominant neutrino heating term
as long as $T_{\nu}\la 10$ MeV we have neglected its contribution 
for the remainder of this paper. For
example, in our reference model the mass loss rate is decreased from 5.1
to $4.9\;10^{13}$ g.cm$^{-2}$.s$^{-1}$ if $\dot{q}_{\nu{\bar \nu}}$ is not 
included (i.e. a reduction of 4.5\%). 
Without $\dot{q}_{\nu{\bar \nu}}$, a non constant neutrino temperature
can be easily implemented. We adopted for $T_{\nu}(r)$ relation (\ref{eq:TThick}) with 
$T_*=2$ MeV and the resulting mass loss rate is then 
$\dot{m}=3.8\;10^{13}$ g.cm$^{-2}$.s$^{-1}$.\\
When $M_\mathrm{BH}$, $T_{\nu}$, $\theta$ or the position $r$ of the foot of 
the line are varied, the mass loss rate behaves as in the optically thin
case 
\begin{eqnarray}
\dot{m}(x) & \approx & 3.8\;10^{13}\mu_\mathrm{BH}\left[\frac{T_{\nu}(x)}{2\ \mathrm{MeV}}\right]^{10} f[x, \theta(x)]\ \ \ \mathrm{g.cm^{-2}.s^{-1}}\nonumber\\
& \approx & 
3.8\;10^{13}\mu_\mathrm{BH}
\left[\frac{T_*}{2\ \mathrm{MeV}}\right]^{10}\times\nonumber\\
& & \left(\frac{r_*}{r}\right)^{15/2}
\left(\frac{1-\sqrt{\frac{r_\mathrm{in}}{r}}}{1-\sqrt{\frac{r_\mathrm{in}}{r_*}}}\right)^{5/2}
f[x, \theta(x)]\ \ \ \mathrm{g.cm^{-2}.s^{-1}}\nonumber\\
\end{eqnarray}
The mass loss rate has been represented in Fig.~\ref{fig:MDotThick} for $T_*=2$ MeV, $r_*=4\
r_h$, $r_\mathrm{in}=3\ r_h$, $r_\mathrm{out}=10\ r_h$
and the same field geometries already considered in Sect.~\ref{sec:NumericalThin}.   
\subsection{Analytical solution}
\label{sec:Analytical}
To obtain an analytical expression for the mass loss rate we have simplified
the original wind problem by making several additional assumptions. We have 
first considered that wind material is not too far from hydrostatic 
equi\-li\-bri\-um
even at the sonic point (in practice, the ratio $v \frac{dv}{dr}/\gamma$
becomes larger than unity at $\frac{y_s-y}{y_s}\sim 5$\%). 
We have also supposed that the pressure is dominated by
the contribution of relativistic particles, which is equivalent to $\eta\ll 1$
since
\begin{equation}
\frac{P_N}{P_r}\simeq 0.5\frac{\eta}{Y_e}
\end{equation}
In practice $\eta\simeq 0.1$ in our numerical solutions, except near the 
disk surface 
but we have nevertheless adopted $P\simeq P_r\propto T^4$ everywhere 
(we take $P_r=\tilde{a} T^4$ with $\tilde{a}$ being given by Eq.~(\ref{eq:Pr}) with $\eta=0.1$).
Finally, we have located the sonic point at $y_\mathrm{s}=y_1$ where
$\gamma(y)=0$.
We then write the temperature 
\begin{equation}
T=T_\mathrm{s}\left(\frac{\rho}{\rho_\mathrm{s}}\right)^{1/3}(1+\tau)
\label{eq:Temp}
\end{equation}
where the $\tau$ function is zero at the sonic point. From the sonic
point down to the disk, as $\eta\propto \rho/T^3$ first remains nearly constant
(see Fig.~\ref{fig:SolutionThin} and \ref{fig:SolutionThick}) $\tau$ stays close to zero. 
It then becomes negative
when $\eta$ increases in the vicinity of the disk.
With these assuptions the wind equations can be rewritten
\begin{equation}
\rho\, v \,s=\dot{m}
\label{eq:FlowApp1}
\end{equation}
\begin{equation}
4\frac{P}{\rho}\;\frac{d\log{T}}{dy}=\gamma r=\frac{d\Phi}{dy}
\label{eq:FlowApp2}
\end{equation}
\begin{equation}
4\frac{P}{\rho}\;\frac{d\log{(1+\tau)}}{dy}=\frac{\dot{q}r}{3 v}=
\frac{\rho s\,\dot{q} r}{3 \dot{m}}
\label{eq:FlowApp3}
\end{equation}
The solution of Eq.~(\ref{eq:FlowApp2}) gives
\begin{equation}
T_0-T=\frac{1}{4 \tilde{a} \Theta} \int_0^y \frac{1}{(1+\tau)^3} \frac{d\Phi}{dy} dy
\end{equation}
where $\Theta=T_\mathrm{s}^3/\rho_\mathrm{s}$ and $T_0$ is the temperature in the
plane of the disk. Taking $\tau=0$ everywhere leads 
to the appro\-xi\-ma\-te solution
\begin{equation}
T_0-T\simeq\frac{1}{4 \tilde{a} \Theta} [\Phi(y)-\Phi_0]
\end{equation}
with $\Phi_0=\Phi(y=0)$. If we moreover suppose that $T_\mathrm{s}\ll T_0$
we get
\begin{equation}
t(y)=\frac{T}{T_0}\simeq \frac{\Phi_\mathrm{s}-\Phi(y)}{\Phi_\mathrm{s}-\Phi_0}
\label{eq:TApp}
\end{equation}
and 
\begin{equation}
\Theta\simeq \frac{\Phi_\mathrm{s}-\Phi_0}{4 \tilde{a} T_0}=\frac{\Delta\Phi}{4 \tilde{a} T_0}
\label{eq:Theta}
\end{equation}
with $\phi_\mathrm{s}=\Phi(y_\mathrm{s})=\Phi(y_1)$.
In Fig.~\ref{fig:SolutionThin} the approximate value of the temperature given by Eq.~(\ref{eq:TApp}) is compared to 
the exact solution obtained in Sect.~\ref{sec:NumericalThin}. It can be seen that the agreement 
is quite satisfactory for $y<1$. 
We then transform Eq.~(\ref{eq:FlowApp3}) using relation (\ref{eq:Temp}), the definition (\ref{eq:TApp}) of $t(y)$ 
and writing $\dot{q}(y)$ as
\begin{equation}
\dot{q}(y)\;=\;Q\;T_0^6\;\left[g(y)-t^6(y)\right]
\end{equation}
where $g(y)=1$ for an optically thin disk with uniform heating. In an 
optically thick disk we only consider the heating due to neutrino 
captures on nucleons and $g(y)$ is the   
geometrical integral appearing in Eq.~(\ref{eq:HeatingNuN}).
We obtain
\begin{equation}
\frac{2}{3}\Theta^2 \;\frac{d(1+\tau)^6}{dy}=\frac{s(y) Q r}{3\dot{m}}\,
T_0^8\,\left\lbrace t^2(y)\,\left[g(y)-t^6(y)\right]\right\rbrace
\end{equation}
Integrating this equation from the disk to the sonic point yields 
\begin{equation}
\dot{m}=\frac{Q r}{2 \tilde{a} \Theta^2[1-(1+\tau_0)^6] }\, T_0^8 \mathcal{I}=
\frac{8 \tilde{a} Q r}{\Delta\Phi^2-h_0^2}\,T_0^{10}\mathcal{I}
\label{eq:MDotApp}
\end{equation}
where the integral
\begin{equation}
\mathcal{I}=\int_0^{y_1}s(y)t^2(y)[g(y)-t^6(y)]\,dy
\end{equation}
can be directly computed from Eq.~(\ref{eq:TApp}), (\ref{eq:Surf}) and (\ref{eq:Potential}).
To obtain Eq.~(\ref{eq:MDotApp}) 
we have used the value (\ref{eq:Theta}) of $\Theta$ and 
\begin{equation}
(1+\tau_0)^3=\frac{T_0}{\rho_0 \Theta}=\frac{4 \tilde{a} T_0^4}{\rho_0 \Delta\Phi}
=\frac{h_0}{\Delta\Phi}
\end{equation}
all quantities with a zero subscript being taken in the plane of the disk.
The analytical formula (\ref{eq:MDotApp}) reproduces the $T_0^{10}$ dependence of $\dot{m}$
and diverges when $h_0=\Delta \Phi$, i.e. when the Bernouilli function
is equal to zero at the disk surface. The mass loss rate is also proportional 
to the mass of the black hole at constant $x=r/r_h$. 
We have tested the accuracy of the analytical expression by computing 
$\dot{m}$ for the reference model with uniform heating (optically thin disk). We get 
$\dot{m}=3.9\,10^{14}$ g.cm$^{-2}$.s$^{-1}$ in reasonable agreement
with the numerical result obtained in Sect.~\ref{sec:NumericalThin}. 
\section{Estimates of the wind Lorentz factor}
\label{sec:Gamma}
In the simple model considered in this paper we do not follow the 
acceleration of the 
wind beyond the sonic point up to relativistic velocities. Therefore, the
terminal Lorentz factor cannot be obtained in a self-consistent way.  
We simply expect that an average value of the terminal Lorentz factor 
will be given by
\begin{equation}
{\bar \Gamma}=\frac{L_\mathrm{w}}{\dot{M}c^2}
\label{eq:Gamma}
\end{equation}
where $L_\mathrm{w}$ is the power injected into the wind  
and 
$\dot{M}$ is the total mass loss from the disk  
\begin{equation}
\dot{M}=\int_{r_\mathrm{in}}^{r_\mathrm{out}}\dot{m}\,2\pi r dr
\end{equation}
The mass loss per unit surface of the disk has been obtained in the 
previous section
\begin{equation}
\dot{m}(x)=\dot{m}_0\, \mu_\mathrm{BH}\left[\frac{T_d(x)}{2\ \mathrm{MeV}}
\right]^{10} f[x, \theta(x)]
\end{equation}
so that
\begin{equation}
\dot{M}=3.4\,10^{12} \,\dot{m}_0 \,\mu_\mathrm{BH}^3\!
\int_{x_\mathrm{in}}^{x_\mathrm{out}}\left[\frac{T_d(x)}{2\ \mathrm{MeV}}
\right]^{10}\!f[x, \theta(x)]\,x dx
\end{equation}
We have computed $\dot{M}$ for $x_\mathrm{in}=3$ and $x_\mathrm{out}=10$, 
both for optically thin and 
optically thick disks.
In optically thin disks the temperature is given by Eq.~(\ref{eq:TThin}) and 
$\dot{m}_0=2.33\,10^{14}$ g.cm$^{-2}$.s$^{-1}$. We have then
\begin{equation}
\dot{M}=5.1\,10^{28}\,
\mu_\mathrm{BH}\left(\frac{\alpha}{0.01}\right)^2
F_\mathrm{geo}
\ \ \ \ \mathrm{g.s}^{-1}
\label{eq:MDotTot}
\end{equation}
with 
\begin{equation}
F_\mathrm{geo}=\int_{x_\mathrm{in}}^{x_\mathrm{out}}\frac{f[x, \theta(x)]}{x^2}\;dx
\end{equation}
For an optically thick disk the temperature of the neutrinosphere is given 
by Eq.~(\ref{eq:TThick}) and $\dot{m}_0=3.8\,10^{13}$ g.cm$^{-2}$.s$^{-1}$ so that
\begin{equation}
\dot{M}=7.9\;10^{28}\,
\mu_\mathrm{BH}^{3}\left(
\frac{T_*}{2\  \mathrm{MeV}}\right)^{10}
F_\mathrm{geo}\ \ \ \ \mathrm{g.s}^{-1}
\end{equation}
with 
\begin{equation}
F_\mathrm{geo}=\int_{x_\mathrm{in}}^{x_\mathrm{out}}\left(\frac{4}{x}\right)^{6.5}
\left(1-\sqrt{\frac{3}{x}}\right)^{5/2}f[x, \theta(x)]\;dx
\end{equation}
We now introduce the ratio $\beta$ of the wind power to the disk neutrino luminosity
\begin{equation}
L_\mathrm{w}=\beta L_{\nu}=\beta \int_{r_\mathrm{in}}^{r_\mathrm{out}}
\frac{7}{8}\sigma T_{\nu}^4(r)\;2\pi r dr
\end{equation}
which yields the following estimate of $T_*$
\begin{equation}
T_*\simeq 2\,\mu_\mathrm{BH}^{-1/2}\left(
\frac{L_{52}}{\beta_5}\right)^{1/4}
\ \ \ \mathrm{MeV}
\label{eq:TStar}
\end{equation}
where $L_{52}=L_\mathrm{w}/10^{52}$ erg.s$^{-1}$ and $\beta_5=\beta/5$.
A large $\beta$ 
corresponds
to a disk able to transfer most of its energy in the Poynting flux 
with limited dissipation and heating.
From Eq.~(\ref{eq:MDotTot}) and (\ref{eq:TStar}) we finally get
\begin{equation}
\dot{M}=7.9\;10^{28}\,
\mu_\mathrm{BH}^{-2}\left(
\frac{L_{52}}{\beta_5}\right)^{5/2}F_\mathrm{geo}\ \ \ \ \mathrm{g.s}^{-1}
\end{equation}

We are now in a position to estimate the average Lorentz factor of the wind
using Eq.~(\ref{eq:Gamma}). The results in the optically thin and optically thick cases 
are given respectively by
\begin{equation}
\begin{array}{rcll}
\bar{\Gamma} & = & 
\frac{220}{F_\mathrm{geo}}\mu_\mathrm{BH}^{-1}\,L_{52}
\left(\frac{\alpha}{0.01}\right)^{-2} & \mathrm{opt.\ thin\ case}\\ 
& & \\
& = & 
\frac{140}{F_\mathrm{geo}}\mu_\mathrm{BH}^{2}\,L_{52}^{-3/2}\,
\beta_5^{5/2}
& \mathrm{opt.\ thick\ case}\\
\end{array}
\label{eq:Estimation}
\end{equation}
and the values of the geometric integrals are listed in Table~\ref{tab:FGeo}.   
\begin{table}
\begin{center}
\begin{tabular}{|l||c|c||c|c|}
\hline
Geometry & \multicolumn{2}{|c||}{Opt. thin disk} &
\multicolumn{2}{|c|}{Opt. thick disk}\\

\hline
& $F_\mathrm{geo}$ & $\bar{\Gamma}$ & $F_\mathrm{geo}$ & $\bar{\Gamma}$\\
\hline
\hline
$\theta=80^{\circ}$ & 9.0  & 24  & 1.1            & 130  \\
\hline
$\theta=85^{\circ}$ & 1.1  & 200 & $9.2\;10^{-2}$ & 1500 \\
\hline
$\theta=89^{\circ}$ & 0.36 & 610 & $2.6\;10^{-2}$ & 5400 \\
\hline
\hline
$\theta=90^{\circ}\!-0.5^{\circ}(x-3)$ & 0.55    & 400   & $4.0\;10^{-2}$ & 3500 \\
\hline 
$\theta=90^{\circ}\!-1.^{\circ}(x-3)$  & 1.3     & 170   & $9.1\;10^{-2}$ & 1500 \\
\hline 
$\theta=90^{\circ}\!-1.5^{\circ}(x-3)$ & $>5.8$  & $<38$ & 0.42           & 340  \\
\hline 
$\theta=90^{\circ}\!-2.^{\circ}(x-3)$  & $>200$  & $\approx 1$  & $>3.3$
 & $<43$\\
\hline 
\end{tabular}
\end{center}
\caption{Values of the geometric integral $F_\mathrm{geo}$ for an
optically thin and an optically thick disk and the field geometries considered 
in Sect. 4. The corresponding values of $\bar{\Gamma}$ are
computed using Eq.~\ref{eq:Estimation} with $\mu_\mathrm{BH}=1$,
$L_{52}=1$ and $\alpha=0.01$ in the optically thin case and $\mu_\mathrm{BH}=1$,
$L_{52}=1$ and $\beta=5$
in the optically thick case.}
\vspace*{-4ex}

\label{tab:FGeo}
\end{table}
\section{Discussion and conclusions}
\label{sec:Discussion}
From Eq.~(\ref{eq:Estimation}) and the values of the geometric integrals in Table~\ref{tab:FGeo} it 
appears that 
the wind emitted from the inner disk can reach large Lorentz factors 
($\Gamma>100$) only under quite restrictive
conditions on the disk temperature and field geometry. 
The disk temperature depends
on the viscosity parameter $\alpha$ (in the optically thin case) 
and on the value of $\beta$
(in the optically thick case). These two quantities which are quite uncertain
unfortunately enter the expression of $\bar{\Gamma}$ with respective powers
2 and 2.5.
The extreme sensitivity of $\bar{\Gamma}$ to the disk temperature also amplifies
all errors and uncertainties in the evaluation of $T_d$.
For example, the simple analytical expression of $T_d$ (Eq. 8) used in the
optically thin case does not depend on the accretion rate 
to the black hole while detailed 
calculations show that $T_d$ slightly increases with $\dot{M}$
\citep{popham:99}. A consequence
of Eq. (8) is that $\bar{\Gamma}\propto L$ but just the opposite behavior
(decreasing $\bar{\Gamma}$ with increasing $L$) would be expected if 
$T_d\propto \dot{M}^x$ with $x>0.1$. It is moreover even not clear that the
$\alpha$-prescription remains appropriate in the context of strongly 
magnetized disks.
\\

The Lorentz factor is also extremely sensitive to the field geometry: 
to escape from the disk, the material has to be heated so that its energy becomes large enough to cross the potential well. The height of this barrier decreases rapidly when the field lines are more inclined, leading to an increasing mass flux. As shown by our analytical solution of the wind equations, there is even a critical inclination where the mass flux diverges. This happens when the potential barrier is so shallow that the initial enthalpy of the material in the disk allows it to freely escape along the field lines, even without additional 
heating. 
For this reason
quasi vertical field 
lines are required to prevent baryonic pollution from growing dramatically and 
slight changes of the inclination angle lead to large variations of
the Lorentz factor. Between $\theta=90$ and 80$^{\circ}$, $\bar{\Gamma}$
approximately behaves as $\theta^{30}$ ! A reduction by 1$^{\circ}$ of 
the inclination angle therefore decreases $\bar{\Gamma}$ by a factor 
of about 1.5.
\\

Our incomplete description of the black hole and its environment
prevents us from making any accurate prediction of the wind Lorentz
factor. The numerical values of the mass loss rate given above have been
obtained with a large set of simplifying assumptions.
We present them to illustrate general tendencies which we believe are robust
and can be used to evaluate the ability of realistic models to produce
relativistic outflows. When the disk is
thick or slim the temperature distribution and the heating or cooling 
processes will be different from the assumptions we have made.
Similarly, the disk + black hole magnetosphere can be expected to have 
a very complex geometry and to be rapidly variable.  
The results of our toy model -- extreme sensitivity of the mass loss rate
to the disk temperature (${\dot m}\propto T_d^{10}$) and field geometry
-- indicate that high terminal Lorentz factors can be reached only under 
severe constraints: disk temperature $T_d$ not largely exceeding 2 MeV and 
presence of at least a few field lines pointing directly away from the disk
in the vertical direction.

If such conditions can be satisfied so that 
the wind can indeed become relativistic, it is then easy
to understand that its baryonic load and hence its Lorentz factor can
strongly vary on short time scales as a result of fluctuations of the
disk temperature or field geometry. Conversely, if the outflow remains
non relativistic ($\bar{\Gamma}\sim 1$) because the disk is too hot 
or the field lines deviate too much from the vertical, the burst must
be produced by the Blandford-Znajek effect alone with no contribution 
from accretion energy. The dense wind emitted from the disk can then have
both a beneficial and negative effect on the central relativistic jet.
It can probably help to confine the jet but can also represent a risk of baryonic
pollution via Kelvin-Helmholtz instabilities or magnetic reconnection
at the jet-wind interface.  
\begin{acknowledgements}
We thank the anonymous referee for careful reading of the manuscript and thoughtful comments.
\end{acknowledgements}

\vspace*{-3ex}

\newcommand{\aap}{A\&A}
\newcommand{\apj}{ApJ}
\newcommand{\apjl}{ApJ}
\newcommand{\apjs}{ApJS}
\newcommand{\mnras}{MNRAS}
\newcommand{\nat}{Nat}
\bibliographystyle{apj}
\bibliography{grbwind}

\begin{thebibliography}{46}
\expandafter\ifx\csname natexlab\endcsname\relax\def\natexlab#1{#1}\fi

\bibitem[{{Artemova} {et~al.}(1996){Artemova}, {Bjoernsson}, \&
  {Novikov}}]{artemova:96}
{Artemova}, I.~V., {Bjoernsson}, G., \& {Novikov}, I.~D. 1996, \apj, 461, 565

\bibitem[{{Baring} \& {Harding}(1997)}]{baring:97}
{Baring}, M.~G. \& {Harding}, A.~K. 1997, \apj, 491, 663

\bibitem[{{Bethe}(1993)}]{bethe:93}
{Bethe}, H.~A. 1993, \apj, 412, 192

\bibitem[{{Bethe} {et~al.}(1980){Bethe}, {Applegate}, \& {Brown}}]{bethe:80}
{Bethe}, H.~A., {Applegate}, J.~H., \& {Brown}, G.~E. 1980, \apj, 241, 343

\bibitem[{{Blandford} \& {McKee}(1976)}]{blandford:76}
{Blandford}, R.~D. \& {McKee}, C.~F. 1976, Physics of Fluids, 19, 1130

\bibitem[{{Blandford} \& {Payne}(1982)}]{blandford:82}
{Blandford}, R.~D. \& {Payne}, D.~G. 1982, \mnras, 199, 883

\bibitem[{{Blandford} \& {Znajek}(1977)}]{blandford:77}
{Blandford}, R.~D. \& {Znajek}, R.~L. 1977, \mnras, 179, 433

\bibitem[{{Calder} {et~al.}(1999){Calder}, {Wang}, \& {Swesty}}]{calder:99}
{Calder}, A.~C., {Wang}, E.~Y.~M., \& {Swesty}, F.~D. 1999, in American
  Astronomical Society Meeting, Vol. 194, 11.602

\bibitem[{{Chevalier} \& {Li}(2000)}]{chevalier:00}
{Chevalier}, R.~A. \& {Li}, Z. 2000, \apj, 536, 195

\bibitem[{{Daigne} \& {Mochkovitch}(1998)}]{daigne:98}
{Daigne}, F. \& {Mochkovitch}, R. 1998, \mnras, 296, 275

\bibitem[{{Daigne} \& {Mochkovitch}(2000)}]{daigne:00}
---. 2000, \aap, 358, 1157

\bibitem[{{Davies} {et~al.}(1994){Davies}, {Benz}, {Piran}, \&
  {Thielemann}}]{davies:94}
{Davies}, M.~B., {Benz}, W., {Piran}, T., \& {Thielemann}, F.~K. 1994, \apj,
  431, 742

\bibitem[{{Dermer} \& {Mitman}(1999)}]{dermer:99}
{Dermer}, C.~D. \& {Mitman}, K.~E. 1999, \apjl, 513, L5

\bibitem[{{Djorgovski et al.}(2001)}]{djorgovski:01}
{Djorgovski et al.} 2001, in Proc. IX Marcel Grossmann Meeting, ed.
  V.~{Gurzadyan}, R.~{Jantzen}, \& R.~{Ruffini} (Singapore: World Scientific),
  in press, ASTROPH 0106574

\bibitem[{{Duncan} {et~al.}(1986){Duncan}, {Shapiro}, \&
  {Wasserman}}]{duncan:86}
{Duncan}, R.~C., {Shapiro}, S.~L., \& {Wasserman}, I. 1986, \apj, 309, 141

\bibitem[{{Faber} {et~al.}(2000){Faber}, {Manor}, \& {Rasio}}]{faber:00}
{Faber}, J.~A., {Manor}, J., \& {Rasio}, F.~A. 2000, in American Astronomical
  Society Meeting, Vol. 196, 38.03

\bibitem[{{Galama et al.}(1998)}]{galama:98}
{Galama et al.} 1998, \nat, 395, 670

\bibitem[{{Hachisu}(1986)}]{hachisu:86}
{Hachisu}, I. 1986, \apjs, 61, 479

\bibitem[{{Herant} {et~al.}(1994){Herant}, {Benz}, {Hix}, {Fryer}, \&
  {Colgate}}]{herant:94}
{Herant}, M., {Benz}, W., {Hix}, W.~R., {Fryer}, C.~L., \& {Colgate}, S.~A.
  1994, \apj, 435, 339

\bibitem[{{Janka} {et~al.}(1999){Janka}, {Eberl}, {Ruffert}, \&
  {Fryer}}]{janka:99}
{Janka}, H.-T., {Eberl}, T., {Ruffert}, M., \& {Fryer}, C.~L. 1999, \apjl, 527,
  L39

\bibitem[{{Jaroszynski}(1993)}]{jaroszynski:93}
{Jaroszynski}, M. 1993, Acta Astronomica, 43, 183

\bibitem[{{Klose et al.}(2000)}]{klose:00}
{Klose et al.} 2000, \apj, 545, 271

\bibitem[{{Kobayashi} {et~al.}(1997){Kobayashi}, {Piran}, \&
  {Sari}}]{kobayashi:97}
{Kobayashi}, S., {Piran}, T., \& {Sari}, R. 1997, \apj, 490, 92

\bibitem[{{MacFadyen} \& {Woosley}(1999)}]{macfadyen:99}
{MacFadyen}, A.~I. \& {Woosley}, S.~E. 1999, \apj, 524, 262

\bibitem[{{Meszaros} \& {Rees}(1992{\natexlab{a}})}]{meszaros:92b}
{Meszaros}, P. \& {Rees}, M.~J. 1992{\natexlab{a}}, \mnras, 257, 29P

\bibitem[{{Meszaros} \& {Rees}(1992{\natexlab{b}})}]{meszaros:92a}
---. 1992{\natexlab{b}}, \apj, 397, 570

\bibitem[{{Meszaros} \& {Rees}(1997)}]{meszaros:97}
---. 1997, \apj, 476, 232

\bibitem[{{Mochkovitch} {et~al.}(1993){Mochkovitch}, {Hernanz}, {Isern}, \&
  {Martin}}]{mochkovitch:93}
{Mochkovitch}, R., {Hernanz}, M., {Isern}, J., \& {Martin}, X. 1993, \nat, 361,
  236

\bibitem[{{Narayan} {et~al.}(1992){Narayan}, {Paczynski}, \&
  {Piran}}]{narayan:92}
{Narayan}, R., {Paczynski}, B., \& {Piran}, T. 1992, \apjl, 395, L83

\bibitem[{{Oohara} \& {Nakamura}(1997)}]{oohara:97}
{Oohara}, K. \& {Nakamura}, T. 1997, in Relativistic Gravitation and
  Gravitational Radiation, 309

\bibitem[{{Ostriker} \& {Mark}(1968)}]{ostriker:68}
{Ostriker}, J.~P. \& {Mark}, J. W.~. 1968, \apj, 151, 1075

\bibitem[{{Owens et al.}(1998)}]{owens:98}
{Owens et al.} 1998, \aap, 339, L37

\bibitem[{{Paczynski}(1998)}]{paczynski:98}
{Paczynski}, B. 1998, \apjl, 494, L45

\bibitem[{{Popham} {et~al.}(1999){Popham}, {Woosley}, \& {Fryer}}]{popham:99}
{Popham}, R., {Woosley}, S.~E., \& {Fryer}, C. 1999, \apj, 518, 356

\bibitem[{{Preece} {et~al.}(1998){Preece}, {Briggs}, {Mallozzi}, {Pendleton},
  {Paciesas}, \& {Band}}]{preece:98}
{Preece}, R.~D., {Briggs}, M.~S., {Mallozzi}, R.~S., {Pendleton}, G.~N.,
  {Paciesas}, W.~S., \& {Band}, D.~L. 1998, \apjl, 506, L23

\bibitem[{{Qian} \& {Woosley}(1996)}]{qian:96}
{Qian}, Y.-Z. \& {Woosley}, S.~E. 1996, \apj, 471, 331

\bibitem[{{Rees} \& {Meszaros}(1994)}]{rees:94}
{Rees}, M.~J. \& {Meszaros}, P. 1994, \apjl, 430, L93

\bibitem[{{Rhoads}(1997)}]{rhoads:97}
{Rhoads}, J. 1997, in Fourth Huntsville Gamma-Ray Burst Symposium, held 15-20
  September, 1997., 02

\bibitem[{{Rosswog} {et~al.}(2000){Rosswog}, {Davies}, {Thielemann}, \&
  {Piran}}]{rosswog:00}
{Rosswog}, S., {Davies}, M.~B., {Thielemann}, F.-K., \& {Piran}, T. 2000, \aap,
  360, 171

\bibitem[{{Ruffert} {et~al.}(1997){Ruffert}, {Janka}, {Takahashi}, \&
  {Schaefer}}]{ruffert:97}
{Ruffert}, M., {Janka}, H.~., {Takahashi}, K., \& {Schaefer}, G. 1997, \aap,
  319, 122

\bibitem[{{Ruffert} \& {Janka}(1999)}]{ruffert:99}
{Ruffert}, M. \& {Janka}, H.-T. 1999, \aap, 344, 573

\bibitem[{{Ruffert} {et~al.}(1996){Ruffert}, {Janka}, \&
  {Schaefer}}]{ruffert:96}
{Ruffert}, M., {Janka}, H.-T., \& {Schaefer}, G. 1996, \aap, 311, 532

\bibitem[{{Sari} {et~al.}(1998){Sari}, {Piran}, \& {Narayan}}]{sari:98}
{Sari}, R., {Piran}, T., \& {Narayan}, R. 1998, \apjl, 497, L17

\bibitem[{{van Paradijs et al.}(1997)}]{vanparadijs:97}
{van Paradijs et al.} 1997, \nat, 386, 686

\bibitem[{{Waxman} {et~al.}(1998){Waxman}, {Kulkarni}, \& {Frail}}]{waxman:98}
{Waxman}, E., {Kulkarni}, S.~R., \& {Frail}, D.~A. 1998, \apj, 497, 288

\bibitem[{{Woosley}(1993)}]{woosley:93}
{Woosley}, S.~E. 1993, \apj, 405, 273

\end{thebibliography}

\end{document}